\documentclass[12pt,a4paper]{article}
\usepackage[DIV13]{typearea}
\usepackage{amsmath,amssymb,amsfonts}
\usepackage{color}
\usepackage[colorlinks]{hyperref}
\usepackage{graphicx} 
\usepackage{axodraw}
\usepackage[small]{caption}
\usepackage[absolute]{textpos}

\newcommand{\GeV}{\:\text{GeV}}
\newcommand{\TeV}{\:\text{TeV}}

\newcommand{\eq}[1]{Eq.~\eqref{#1}}
\newcommand{\Figref}[1]{Fig.~\ref{#1}}
\newcommand{\Tabref}[1]{Tab.~\ref{#1}}
\newcommand{\Secref}[1]{Sec.~\ref{#1}}

\newcommand{\nn}{\nonumber}
\def\bea{\begin{eqnarray} }
\def\eea{ \end{eqnarray} } 

\newcommand{ \Cpm       }{\tilde{\chi}^\pm}
\newcommand{ \neut      }{\tilde{\chi}^0}
\newcommand{ \Mch       }{M_{\Cpm}}
\newcommand{ \Mneut     }{M_{\neut}}
\newcommand{ \cv        }{\cos\theta_{\tilde{\nu}}}
\newcommand{ \sv        }{\sin\theta_{\tilde{\nu}}}

\newcommand{ \twosv     }{\sin^2 \theta_{\tilde{\nu}}}
\newcommand{ \twocv     }{\cos^2 \theta_{\tilde{\nu}}}

\newcommand{ \snu       }{\tilde{\nu}}

\newcommand{ \deltaLL   }{(\delta_{12}^l)_{LL}^{}}
\newcommand{ \deltaRR   }{(\delta_{12}^l)_{RR}^{}}
\newcommand{ \FVAbs     }{|m_{\tilde L_{12}}^2|}

\DeclareMathOperator{\re}{Re}
\DeclareMathOperator{\diag}{diag}
\newcommand{\maxx}{\text{EXP}}

\definecolor{darkblue}{rgb}{0,  0,  .5}
\definecolor{lightgray}{gray}{0.5}

\newcommand{\lightgray}[1]{\textcolor{lightgray}{#1}}

\newcommand{\ramTL}{\frac{\amuegL}{a_\mu}}
\newcommand{\ramTR}{\frac{\amuegR}{a_\mu}}
\newcommand{\ramTLS}{\amuegL/a_\mu}
\newcommand{\ramTRS}{\amuegR/a_\mu}
\newcommand{\rCh}{\frac{a^{\tilde \chi^{\pm}_k}_{\mu e \gamma\ I}} {a^{\tilde \chi^\pm_k }_{\mu\ I} } }
\newcommand{\rChS}{{a^{\tilde \chi^{\pm}_k}_{\mu e \gamma\ I}} /{a^{\tilde \chi^\pm_k }_{\mu\ I} } }
\newcommand{\rChSII}{{a^{\tilde \chi^{\pm}_k}_{\mu e \gamma\ 2}} /{a^{\tilde \chi^\pm_k }_{\mu\ 2} } }

\newcommand{\amuII}{a^{\tilde \chi^\pm_k }_{\mu\ 2} }
\newcommand{\amuegII}{a^{\tilde \chi^{\pm}_k}_{\mu e \gamma\ 2}}
\newcommand{\amuegI}{a^{\tilde \chi^{\pm}_k}_{\mu e \gamma\ 1}}
\newcommand{\amuegG}{a_{\mu e \gamma}}
\newcommand{\amuG}{a_{\mu}}
\newcommand{\amuegL}{a_{\mu e \gamma L}}
\newcommand{\amuegR}{a_{\mu e \gamma R}}
\newcommand{\BR}{\ensuremath{\text{BR}(\mu \to e \gamma)}}

\begin{document}

\begin{titlepage}

\begin{center}
{\Large\bf\boldmath 
Understanding the correlation between $(g-2)_\mu$ and $\mu \rightarrow e \gamma$ in the MSSM\\
}

\vspace{1cm}
\renewcommand{\thefootnote}{\arabic{footnote}}

\textbf{
J\"orn Kersten\footnote[1]{Email: \texttt{joern.kersten@desy.de}}$^{(a,b)}$, 
Jae-hyeon Park\footnote[2] {Email: \texttt{jae.park@uv.es}}$^{(c)}$,
Dominik St\"ockinger\footnote[3]{Email: \texttt{Dominik.Stoeckinger@tu-dresden.de}}$^{(d)}$\\
and
Liliana Velasco-Sevilla\footnote[4]{Email: 
\texttt{liliana.velascosevilla@gmail.com}}$^{(a)}$
}
\\[5mm]
\textit{\small 
$^{(a)}$
University of Hamburg, II.\ Institute for Theoretical Physics,\\
Luruper Chaussee 149, 22761 Hamburg, Germany\\[2mm]
$^{(b)}$
University of Bergen, Department of Physics and Technology,\\
PO Box 7803, 5020 Bergen, Norway\\[2mm]
$^{(c)}$
Departament de F\'{i}sica Te\`{o}rica and IFIC,\\
Universitat de Val\`{e}ncia-CSIC,
46100, Burjassot, Spain\\[2mm]
$^{(d)}$Institut f\"ur Kern- und Teilchenphysik, TU Dresden, 01069 Dresden, Germany
\\[2mm]
}
\end{center}

\vspace{1cm}

\begin{abstract}
\noindent
The supersymmetric contributions to the muon anomalous magnetic moment $a_\mu$ and to the
decay $\mu\to e\gamma$ are given by very similar Feynman
diagrams.  Previous works reported correlations in specific scenarios, in particular if
$a_\mu$ is dominated by a single diagram. In this work we give an extensive survey of the possible correlations.  We discuss examples of single-diagram domination with particularly
strong correlations, and provide corresponding benchmark parameter
points. We show how the correlations are weakened by significant
cancellations between diagrams in large parts of the MSSM parameter
space. Nevertheless, the order of magnitude of $\BR$ for a fixed
flavor-violating parameter can often be predicted. We summarize the behavior by plotting the correlations as well as resulting bounds on the flavor-violating parameters under various assumptions on the MSSM spectrum.
\end{abstract}

% report numbers
\begin{textblock*}{7em}(\textwidth,1cm)
\noindent\footnotesize
FTUV--14--1261 \\
IFIC--14--39
\end{textblock*}

\end{titlepage}

\section{Introduction}
 
As is well established, the Minimal Supersymmetric Standard Model (MSSM)
with fully general soft supersymmetry (SUSY) breaking parameters is
strongly constrained by flavor-violating observables. Particularly in
the charged lepton sector no sign of flavor violation has been
observed yet. In contrast, the muon anomalous magnetic moment
$a_\mu = (g_\mu-2)/2$ is
a flavor-conserving leptonic observable with a tantalizing deviation
$\Delta a_\mu$ between experiment and Standard Model (SM) prediction by more
than $3\sigma$, which could be explained beautifully by the
contributions from light sleptons, charginos, or neutralinos.
On the other hand, the Feynman diagrams for SUSY contributions to $a_\mu$ and to the
branching ratio of the flavor-violating decay $\mu\to e\gamma$, \BR, are
essentially identical, except for the flavor transition appearing only in the latter case. For this reason it was put forward to study correlations between the two observables in the MSSM
\cite{Graesser:2001ec,Chacko:2001xd,Isidori:2007jw}.

Such correlations could prove useful, for example, in constraining scenarios which
are in agreement with $a_\mu$. Once the SUSY contribution to $a_\mu$ is fixed, the mass scale of the
relevant superparticles is fixed as well.  One could then predict the
value of \BR\ as a function of only the flavor-violating SUSY breaking
parameters and thus derive stringent bounds on these parameters, which
cannot be evaded by simply raising the overall SUSY mass scale.
It has already been noted in \cite{Chacko:2001xd} that a strong correlation could emerge
if the contribution to $a_\mu$ comes mainly from a single diagram; a strong correlation was also observed
in \cite{Isidori:2007jw}  in a parameter scan.
The interest of the present work is to identify the parameter space
where this happens.  In fact, cases with a strong correlation often correspond
to certain mass hierarchies among the particles involved. Hence we characterize such hierarchies and establish the corresponding bounds.

On the other hand, in large parts of the MSSM parameter space the SUSY
contribution to $a_\mu$ is large but no single contribution to $a_\mu$
dominates. It is then also of interest to study to what extent the two
observables are correlated and whether we can derive bounds on the
flavor-violating parameters even in such cases.

Motivated by the stringent LHC mass limits on colored SUSY particles,
which in constrained scenarios imply stringent limits also on uncolored
superparticles, we consider the 
 MSSM without assumptions on
GUT-scale or SUSY breaking physics.
For our study, only electroweak parameters are relevant: the higgsino and gaugino mass parameters $\mu$, $M_1$, $M_2$, the ratio of the Higgs vacuum expectation values $\tan\beta$, and the slepton mass and mixing parameters, to be described below.

The work is organized as follows. In \Secref{sec:ExpStatus} we present
the status of the relevant observables. In \Secref{sec:Contributions} we
discuss the SUSY contributions to $a_\mu$ and \BR. In
\Secref{sec:SimilarMasses}, we survey their correlation in the general
case where all supersymmetric particles involved in $a_\mu$ and
$\mu \to e\gamma$ have no particular hierarchy. For this case the
charginos tend to dominate, so in \Secref{sec:chargino dominance} we
study the conditions under which correlations between $a_\mu$ and $\BR$
could be established for chargino domination. In \Secref{sec:LargeMu} we
study a scenario where the $\mu$ parameter is very large and the
lightest neutralino is essentially a bino, so the main contribution to
$a_\mu$ is given by diagrams involving binos, $\tilde\mu_L$, and
$\tilde\mu_R$.  Finally, in \Secref{sbsc:neutsmuR} we consider the case
where all left-handed sleptons are very heavy, so that the chargino and
most neutralino contributions are suppressed, except for the neutralino
contribution with $\tilde\mu_R$ exchange.  

Section~\ref{sec:discussions} can be read independently. It provides an
extended discussion of the results, and it summarizes the behavior with
plots of the correlations and bounds on flavor-violating parameters
under various assumptions on the SUSY spectrum.

\section{Status of relevant observables} \label{sec:ExpStatus}

\subsection{$\mathbf{(g-2)_\mu}$ and BR$(\mu\rightarrow e\gamma)$}
The difference between the experimental determination \cite{Bennett:2006}
and the SM prediction for the anomalous magnetic moment of
the muon is larger than $3\sigma$. Taking the evaluation of hadronic
contributions of  Ref.\
\cite{Davier:2010nc}, including recent updates of the QED
\cite{Kinoshita2012} and electroweak \cite{Gnendiger:2013pva}
contributions, and adding theoretical and experimental uncertainties
in quadrature, the difference is
\begin{equation}
\label{eq:Dmuallw_exp}
\Delta a_\mu = a_\mu^\text{exp} - a_\mu^\text{SM} = (287\pm80)\times 10^{-11}.
\end{equation}
Alternative
theory evaluations \cite{Hagiwara:2011af,Benayoun:2012wc} obtain similar
or even larger
 differences. Further progress can be expected not only from
 improvements on the theory side, but in particular from  Fermilab P989
 \cite{Carey:2009zzb,Roberts:2010cj}  and the new J-PARC approach to the
 $g-2$/EDM measurements \cite{Iinuma:2011zz}. Both aim to
 improve the experimental uncertainty of $a_\mu$ by at least a
 factor of 4.

 In quantum field
 theory, $a_\mu$ can be obtained from the covariant decomposition of
 the muon--photon three-point function. Written similarly
 to \cite{Chacko:2001xd}, the relevant term is
\begin{equation} \label{eq:Amplitudeg-2}
\mathcal{M}_\mu= \frac{e}{2m_\mu} \, \epsilon^\alpha \bar{u}_\mu(k+q)
\, [ i q^\beta \sigma_{\beta\alpha}  a_\mu ] \, u_\mu(k),
\end{equation}
in the limit $q\to0$. Here
$\sigma_{\alpha\beta} = i/2 \, [\gamma_\alpha,\gamma_\beta]$,
$\epsilon^\alpha$ is the
photon polarization vector, $k$ and $k+q$ are on-shell momenta, and finally
$u_\mu$, $\bar u_\mu$ are spinors that satisfy the Dirac equation. 

The current 90\% C.L.\ upper limit on the branching ratio
$\text{BR}(\mu\rightarrow e\gamma)$, set by the MEG experiment
\cite{Adam:2013mnn}, is
\bea
\label{expval:BRemugamma}
\text{BR}(\mu\rightarrow e\gamma) < 5.7 \times 10^{-13} \equiv
\text{BR}_{\maxx}(\mu\rightarrow e\gamma),
\eea
and future upgrades  will attempt to explore  regions of $O(10^{-14})$ \cite{Baldini:2013ke}. We write the amplitude for $\mu\rightarrow e \gamma $ as
\begin{equation} \label{eq:Amplitudemueg}
\mathcal{M}_{\mu e \gamma} = \frac{e}{2m_\mu}\, \epsilon^{*\alpha}
\bar{u}_e(k+q)
\,[ i \sigma_{\beta\alpha} q^\beta
( a_{\mu e \gamma R} P_L + a_{\mu e \gamma L} P_R) ] \, u_\mu (k) ,
\end{equation}
where $P_{R,L}=(1\pm \gamma_5)/2$. The $L/R$ index in
$a_{\mu e \gamma L/R}$ refers to the electron chirality.
Thus, $L$ and $R$ are interchanged with respect to the notation in
\cite{Hisano:1995cp,Chacko:2001xd}.
By convention, the photon momentum $q$ is oriented towards the vertex
in both \eq{eq:Amplitudeg-2} and \eq{eq:Amplitudemueg}. The resulting branching ratio is
\begin{equation}
\label{def:BRemugamma}
\text{BR} (\mu\rightarrow e \gamma)= \frac{3\pi^2 e^2}{G_F^2 m^4_\mu} \, ( |a_{\mu e \gamma L}|^2 + |a_{\mu e \gamma R}|^2 ).
\end{equation}

\subsection{Superparticle masses}

While the negative results from LHC SUSY searches place stringent lower
limits on the masses of the colored superparticles, the electroweakly
interacting charginos, neutralinos and sleptons are still allowed to be
quite light
\cite{Fowlie:2013oua,Cahill-Rowley:2013yla,Henrot-Versille:2013yma,Calibbi:2013poa,Belanger:2013pna}.
This is very enthralling because a sizable supersymmetric contribution
to the muon $g-2$ requires masses of $\mathcal{O}(100)\GeV$ precisely for sleptons
and charginos or neutralinos.  Consequently, within the general MSSM an
appealing possibility is to consider scenarios with the hierarchy
\begin{equation}
m_{\tilde q}, \; m_{\tilde g} \gg m_{\tilde \ell }, \; m_{\Cpm}, \; m_{\neut}.
\label{masshierarchy}
\end{equation}

Collider constraints on the masses relevant for $a_\mu$ and $\mu\to e\gamma$ for certain specific scenarios of this kind
were derived in \cite{Endo:2013bba,Fowlie:2013oua}.
In particular in \cite{Endo:2013bba} the recent
results on the searches for the non-colored supersymmetric particles
were investigated in the parameter region where the muon $g-2$ is
explained. Under the assumption of the GUT relation $M_1\approx
M_2/2$, lower bounds of around $150$--$200\GeV$ were obtained for the
wino mass.
Further, Refs.\ \cite{Fargnoli:2013zda,Fargnoli:2013zia} revealed
logarithmically enhanced two-loop corrections from mass hierarchies
such as those in \eq{masshierarchy}. Since we will
focus on the parametric dependence 
of $a_\mu$, $\BR$ and their correlation on
the SUSY spectrum, we will choose parameters 
satisfying \eq{masshierarchy} but we
will not use more detailed LHC mass limits and restrict ourselves to
one-loop accuracy.

\section[Contributions to amu and amuegamma]{Contributions to $\boldsymbol{a_\mu}$ and $\boldsymbol{a_{\mu e \gamma}}$}
\label{sec:Contributions}

\subsection{Chargino--sneutrino contributions}

The one-loop contributions to $g-2$ in the MSSM have been evaluated in full
generality in \cite{Moroi:1995yh}.
If the mixing of sneutrinos of the first two generations can be
decoupled from the mixing of the third generation, the 
contributions to $g-2$ from a chargino $\tilde\chi^\pm_k$ can be
nicely decomposed into two terms 
\begin{equation} \label{eq:Splitamu}
a_\mu^{\tilde\chi^\pm_k} = a_\mu^{\tilde\chi^\pm_k}{}_1+a_\mu^{\tilde\chi^\pm_k}{}_2
\end{equation}
with
\begin{align}
\label{eq:chargsneutcont}
\frac{16\pi^2}{m_\mu} \, a_\mu^{\tilde\chi^\pm_k}{}_{1} &= \frac{m_\mu}{12 m_{\Cpm_k}^2}
   \left( g_2^2 |V_{k1}|^2+    Y_\mu^2 |U_{k2}|^2\right) \left[ \twosv \, x_{k1} F_1^C(x_{k1}) 
    + \twocv \, x_{k2} F_1^C(x_{k2}) \right],
\nonumber \\
\frac{16\pi^2}{m_\mu} \, a_\mu^{\tilde\chi^\pm_k}{}_{2} &= -\frac{2}{3 m_{\Cpm_k}}
    \, g_2 Y_\mu \re[V_{k1} U_{k2}] \left[ \twosv \, x_{k1} F_2^C(x_{k1}) 
    + \twocv \, x_{k2} F_2^C(x_{k2}) \right],
\end{align}
which correspond to the diagrams mediated by winos or
higgsinos, and a combination of higgsino and wino, respectively, and
which involve corresponding powers of the gauge and muon Yukawa
couplings $g_2$ and $Y_\mu$.
Throughout this work we use the conventions of
\cite{Stockinger:2006zn} unless specified otherwise.
For completeness, the well-known functions $F_1^C$ and $F_2^C$ are defined as
\begin{align}
\label{eq:loopfunctsCha}
F_1^C(x) &= \frac{2}{(1 - x)^4} 
             \left[ 2 + 3 x - 6 x^2 + x^3 + 6 x \ln x \right], \nn\\
F_2^C(x) &= -\frac{3}{2 (1 - x)^3} 
             \left[ 3 - 4 x + x^2 + 2 \ln x \right] ;
\end{align}
the arguments $x_{ki}$ are defined as the mass ratios 
\bea
\label{eq:defratioXki}
x_{ki}\equiv\frac{m^2_{\Cpm_k}}{m_{\tilde\nu_i}^2} .
\eea
The sneutrino mass eigenvalues $m_{\tilde\nu_i}$ and the mixing angle
between the first two generations, $\theta_{\tilde\nu}$, are defined
via diagonalization of the sneutrino mass matrix,
\begin{align}
\begin{pmatrix} \cv & \sv \\ -\sv & \cv \end{pmatrix}
\begin{pmatrix}
m_{\tilde L_{11}}^2 + \mathcal{D}^\nu_L & m_{\tilde L_{12}}^2 \\ 
m_{\tilde L_{12}}^2 & m_{\tilde L_{22}}^2 + \mathcal{D}^\nu_L
\end{pmatrix}
\begin{pmatrix} \cv & -\sv \\ \sv & \cv \end{pmatrix}
&= \diag(m_{\tilde\nu_1}^2,m_{\tilde\nu_2}^2)
\nonumber\\
\Rightarrow\quad
\tan 2 \theta_{\tilde\nu}&=\frac{2 m_{\tilde L_{12}}^2 }{m_{\tilde L_{11}}^2 - m_{\tilde L_{22}}^2},
\label{eq:SneutrinoMixing}
\end{align}
and the requirement that $\snu_1$ be the mass eigenstate composed
primarily of $\snu_e$, which implies that $m_{\snu_1}$ is not
necessarily smaller than $m_{\snu_2}$. Here
$m^2_{\tilde L_{11}}$, $m^2_{\tilde L_{22}}$, $m_{\tilde L_{12}}^2$ are the
flavor-diagonal and off-diagonal soft mass parameters, and ${\cal D}_L^\nu$
are the $D$-term contributions to the masses.
For the chargino mass matrix $\Mch$
we use the diagonalization
\begin{equation}
\label{eq:Chmixconv}
U^* \Mch V^{\dagger} = \diag(m_{\Cpm_1},m_{\Cpm_2}) .
\end{equation}

In the following we will neglect terms suppressed by two powers of the
muon Yukawa coupling or by the electron Yukawa coupling. Then, as
pointed out in \cite{Chacko:2001xd}, the right-handed part of the
$\mu\to e\gamma$ amplitude vanishes,
\begin{equation}
\amuegR^{\tilde\chi^\pm_k} = 0.
\end{equation}
And for the left-handed part we
can write
\begin{equation} \label{eq:SplitamuegL}
\amuegL^{\tilde\chi^\pm_k} = \amuegI + \amuegII ,
\end{equation}
where for each chargino and for each 
of the two contributions in Eqs.~\eqref{eq:Splitamu} and
\eqref{eq:SplitamuegL}, the ratio of the
left-handed $\mu \to e\gamma$ amplitude and $a_\mu$ takes the compact form
\begin{equation}
\label{eq:corr_mueg_mu}
\rCh=\frac{\sin 2\theta_{\tilde\nu}}{2}\frac{x_{k1} F_I^C(x_{k1}) - x_{k2} F_I^C(x_{k2}) }
{ \sin^2 \theta_{\tilde\nu} \, x_{k1} F_I^C(x_{k1}) + \cos^2 \theta_{\tilde \nu} \, x_{k2} F_I^C(x_{k2}) }
%\quad
\ ,\quad I=1,2 .
\end{equation}
We can rewrite this ratio as
\begin{equation}
\label{eq:corr_mueg_mu2}
\rCh=\frac{m_{\tilde L_{12}}^2}{m_{\tilde\nu_1}^2-m_{\tilde\nu_2}^2} \,
\frac{\Delta_I}{x_{k_2} F_I^C(x_{k2}) + \sin^2\theta_{\tilde\nu} \Delta_I}\ ,
\end{equation}
where $\Delta_I \equiv x_{k1} F_I^C(x_{k1}) - x_{k2} F_I^C(x_{k2})$.
We will use this result to derive analytical approximations later on.

\subsection{Neutralino--charged slepton contributions}

The contributions from the neutralinos and charged sleptons
can be decomposed similarly to the chargino--sneutrino contributions,
but the mixing structure is more complicated. Even if only the first
two generations are allowed to mix, the relevant mixing of the left- and
right-handed charged sleptons is described by a $4\times4$ matrix. In
\cite{Chacko:2001xd}, this matrix was diagonalized approximately.
In the following
analysis, we consider the full mixing structure. Then, the neutralino
contributions to  $a_{\mu}$ can be written as (compare Ref.\ 
\cite{Stockinger:2006zn} for the flavor-diagonal result)
\begin{align}
\frac{16\pi^2}{m_\mu} \, a_{\mu}^{\tilde\chi^0_i} =
\sum_{m} \bigg[&
-\frac{m_\mu}{12 m^2_{\tilde\chi_i^0}}
\left[n^{L*}_{\mu im}n^L_{\mu im} + n^{R}_{\mu im}n^{R*}_{\mu im}\right]
x_{im} F_1^N(x_{im}) 
\nonumber\\
& {} +
\frac{1}{3 m_{\tilde\chi_i^0}} \re[n^{L}_{\mu im} n^{R}_{\mu im}] \,
x_{im} F_2^N(x_{im}) \bigg].
\end{align}
Neglecting Yukawa-suppressed terms like in the chargino case, the
$\mu\to e\gamma$ amplitudes can be written as
\begin{align}
\frac{16\pi^2}{m_\mu} \, \amuegR^{\tilde\chi^0_i} & =
\sum_{m} \biggl[ -\frac{m_\mu}{12 m^2_{\tilde\chi_i^0}}
\, n^{R}_{\mu im} n^{R*}_{eim} \, x_{im} F_1^N(x_{im}) +
\frac{1}{3 m_{\tilde\chi_i^0}} \, n^{L*}_{\mu im} n^{R*}_{eim} \, x_{im} F_2^N(x_{im})
\biggr] ,
\nn\\
\frac{16\pi^2}{m_\mu} \, \amuegL^{\tilde\chi^0_i} & =
\sum_{m} \biggl[ -\frac{m_\mu}{12 m^2_{\tilde\chi_i^0}}
\, n^{L*}_{\mu im}n^{L}_{eim} \, x_{im} F_1^N(x_{im}) +
\frac{1}{3 m_{\tilde\chi_i^0}} \, n^{R}_{\mu im} n^{L}_{eim} \, x_{im} F_2^N(x_{im})
\biggr].
\end{align}
In an obvious analogy to the chargino case, one could introduce
$a_{\mu}^{\tilde\chi^0_i}{}_{1,2}$ etc., but we will not make
use of that.
In the previous equations, the abbreviations are defined as
\begin{align}
n^L_{\ell im} &= \frac{1}{\sqrt{2}} \left( g_1 N_{i1} + g_2 N_{i2} \right) K^*_{m,\ell}
- Y_\ell N_{i3} K_{m,\ell+3}^* ,
\nonumber\\
n^R_{\ell im} &=
\sqrt{2} g_1 N_{i1} K_{m,\ell+3} +
Y_\ell N_{i3} K_{m,\ell} ,
\end{align}
the loop functions are
\begin{align}
F_1^N(x) &= \frac{2}{(1 - x)^4} 
             \left[ 1 - 6 x + 3 x^2 + 2 x^3 - 6 x^2 \ln x \right], \nn\\
F_2^N(x) &= \frac{3}{(1 - x)^3} 
             \left[ 1 - x^2 + 2 x \ln x \right] ,
\end{align}
and
\begin{equation}
\label{eq:defratioXim}
x_{im}\equiv\frac{m^2_{\tilde\chi^0_i}}{m_{\tilde\ell_m}^2} .
\end{equation}
To define the slepton masses and mixing we start from the slepton mass
terms for the interaction and flavor eigenstates $\tilde{\ell}_{Li}$,
$\tilde{\ell}_{Ri}$ with generation index $i=1,2,3$,
\begin{equation}
(\tilde \ell^*_{L1}, \tilde \ell^*_{L2},\hdots,\tilde \ell^*_{R3})
\, \mathcal{M}^2
\begin{pmatrix}
\tilde \ell_{L1}\\
\tilde \ell_{L2}\\
\vdots\\
\tilde \ell_{R3}\\
\end{pmatrix} =
(\tilde \ell^*_{L1}, \tilde \ell^*_{L2},\hdots,\tilde \ell^*_{R3})
\begin{pmatrix}
m^2_{\tilde L} + \mathcal{D}^\ell_L & m^{2\dagger}_{LR}\\
m^{2}_{LR} & m^2_{\tilde R} + \mathcal{D}^\ell_R 
\end{pmatrix}
\begin{pmatrix}
\tilde \ell_{L1}\\
\tilde \ell_{L2}\\
\vdots\\
\tilde \ell_{R3}\\
\end{pmatrix}
\label{eq:mass_flav_egs}
\end{equation}
with $3\times3$ block matrices $m^2_{\tilde L}$, $m^2_{\tilde R}$ and
$(m^2_{LR})_{ij} = \delta_{ij} m_i (A_i-\mu^*\tan\beta)$.
We have omitted the small $F$-term contribution $m_i^2$ involving the lepton
masses.  We use a basis where the lepton mass matrix is diagonal.
The $6\times6$ mass matrix $\mathcal{M}^2$ leads to slepton mass
eigenvalues $m_{\tilde\ell_1} < m_{\tilde\ell_2} < 
\hdots < m_{\tilde\ell_6}$ and the diagonalization matrix
$K_{m\ell}$ defined via
\begin{equation}
\label{eq:diagmatKl}
K \mathcal{M}^2 K^{\dagger} =
\diag(m_{\tilde\ell_1}^2,\hdots,m_{\tilde\ell_6}^2).
\end{equation}
For the neutralino mass and mixing matrices
we use the convention
\begin{equation}
\label{eq:neutrinodiag}
N^* \Mneut N^{-1} = \diag(m_{\tilde \chi^0_1},\hdots, m_{\tilde \chi^0_4}).
\end{equation}

\section{Correlations in different parameter space regions \label{eq:scenarios}}

\subsection{Similar SUSY masses \label{sec:SimilarMasses}}

\begin{figure}
\centering
\includegraphics[scale=0.95]{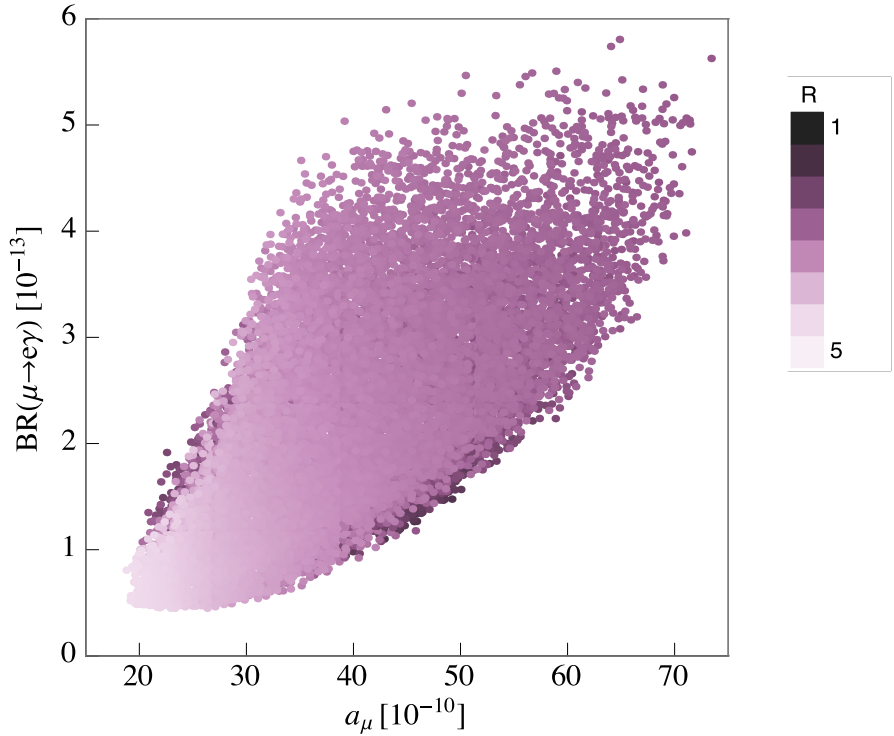}
\caption{Scatter plot ($10^5$ points) of $a_\mu$ versus \BR\ for
 similar SUSY mass parameters between $300$ and $600$ GeV, $\tan\beta=50$,
 and $\deltaLL = \deltaRR = 2 \times 10^{-5}$.
 The points are color-coded according to the degree of cancellation $R$, cf.\ \eq{eq:DefR}. }
\label{fig:ScatterEqualM-R}
\end{figure}

In order to obtain a first impression of the strength of the correlation
between the SUSY contribution
\begin{equation}
a_\mu \equiv \sum_k a_\mu^{\tilde\chi^{\pm}_k} + \sum_i a_\mu^{\tilde\chi^0_i}
\end{equation}
and \BR\ for generic SUSY spectra without strong mass
hierarchies between sleptons, neutralinos and charginos, we
perform a random scan over the parameters
\[
	M_1 ,\; M_2 ,\; \mu ,\; m_{\tilde L_{11}} ,\; m_{\tilde L_{22}} ,\;
	m_{\tilde R_{11}} ,\; m_{\tilde R_{22}} \;,
\]
varying them between $300$ GeV and $600$ GeV while ensuring that a
neutralino is the lightest superparticle (LSP).  
All other superparticle masses are irrelevant for our calculations.
We set the trilinear couplings $A_e$ and $A_\mu$ to zero after verifying
that they have no impact.  We fix $\tan\beta=50$ and
$\deltaLL = \deltaRR = 2 \times 10^{-5}$ for the flavor-violating
parameters
\begin{equation} \label{eq:classicfvp}
\deltaLL \equiv \frac{m^2_{\tilde L_{12}}}{\sqrt{m^2_{\tilde L_{11}} m^2_{\tilde L_{22}}}}
\quad,\quad
\deltaRR \equiv \frac{m^2_{\tilde R_{12}}}{\sqrt{m^2_{\tilde R_{11}} m^2_{\tilde R_{22}}}} .
\end{equation}
The results are shown in \Figref{fig:ScatterEqualM-R}.
Taking into account that the correlated quantities are the amplitudes
$a_\mu$, $\amuegL$ and $\amuegR$ whereas $\BR$ involves amplitudes
squared, we observe a correlation that is significant but not extremely
strong. This indicates that typically several diagrams contribute,
either cancelling each other or adding up constructively.  In order to
investigate this, we encoded in the color of points the importance
of the leading contribution, defined as
\begin{equation} \label{eq:DefR}
	R \equiv
	 \frac{ \max_{i,k}\bigl\{ |a_\mu^{\tilde\chi^{\pm}_k}|,|a_\mu^{\tilde\chi^0_i}| \bigr\}}{a_\mu} .
\end{equation}
$R$ is a measure for the degree of cancellation:
If a single diagram dominates, $R \approx 1$, while $R>1$ indicates
cancellations; $R<1$ if diagrams add up constructively.  The
figure shows that the latter does not occur.  In fact, $R$ is larger
than about $1.5$ for all points.  For more than $90\%$ of them we find
even $R>2$.  Consequently, significant cancellations are typical.

We have checked that adding the constraints $m^2_{\tilde L_{11}} = m^2_{\tilde L_{22}}$
or $m^2_{\tilde R_{11}} = m^2_{\tilde R_{22}}$ does not strengthen the
correlation significantly.
It is also virtually unchanged if we set $\deltaRR=0$ while keeping
$\deltaLL = 2 \times 10^{-5}$.  On the other hand, the correlation
becomes very weak if only $\deltaRR$ is non-zero.  This is to be
expected, since only neutralinos contribute to $\amuegG$ for
$\deltaLL=0$, whereas $a_\mu$ is dominated by charginos
for similar SUSY masses.
Finally, the correlation is not very sensitive to the value of
$\tan\beta$, since the dominant contributions to all amplitudes are
proportional to this parameter, so it does not appear in their ratios,
see e.g.~\eq{eq:corr_mueg_mu}.
Exceptions can occur for $\tan\beta<10$, but in this case $a_\mu$ is
generically too small to be of interest.

In the following we will study scenarios for which the
correlation can become stronger.  They are characterized by hierarchies
within the SUSY spectrum, and \Tabref{tbl:ex_points} already shows the
benchmark parameter choices considered in sections~\ref{sec:chargino dominance}, \ref{sec:LargeMu} and \ref{sbsc:neutsmuR}.  Comparing these cases with the generic,
non-hierarchical case discussed here will further clarify the weak
correlation in  \Figref{fig:ScatterEqualM-R}. Later, in
\Secref{sec:discussions}, we will compare 
bounds on the flavor-violating parameters for all these cases.

\begin{table}
\centering
\begin{tabular}{|c|c|c|c|}
\hline
Parameter $/$ Kind of spectrum & II   & III &IV \\
\hline\hline
$\tan\beta$              &    50                               &  50                    & 50   		 \\
$\mu$                       &   500                & 150\,\ldots\,5000      & $-550$\,\ldots\,$-650$      \\
$M_2$                      &  550\,\ldots\,1300                    & 1800  	 	 &   100\,\ldots\,900   \\
$M_1$                      &   \lightgray{400}                   &  320 	           &   100     \\
$m_{\tilde L_{12}}$ &               3 	    	           &2  			 & 0\\
$m_{\tilde L_{11}}$ & \lightgray{502}           &  470 		 & \lightgray{3001} 
\\
$m_{\tilde L_{22}}$ & \lightgray{502}	          & 490        	&  \lightgray{3001} 	
\\
$m_{\tilde R_{12}}$ &   0                                &  0   		         &  2 \\
$m_{\tilde R_{11}}$ & \lightgray{900}      	          & 510 		& 120      \\
$m_{\tilde R_{22}}$ & \lightgray{950}                  & 600	          & 80\,\ldots\,250   \\
$m_{\tilde\nu_1}$    & 496                           & \lightgray{463}      & \lightgray{3000}  
 \\
$m_{\tilde\nu_2}$    & 496                           & \lightgray{483}       & \lightgray{3000}   \\
\hline
\end{tabular}
\caption{%
Benchmark parameter choices II (chargino dominance with similar
masses, \Secref{sec:chargino dominance}),
III (large $\mu$, \Secref{sec:LargeMu}), and
IV (neutralino--$\tilde{\mu}_R$ dominance, \Secref{sbsc:neutsmuR}).
Masses are given in GeV\@.
Numbers printed in gray denote parameters that do not have an important
impact on the values of $a_\mu$ and $\BR$ for the respective benchmark. Note that the sneutrino masses are not independent of the other input parameters, and their given values are rounded to integer values.
\label{tbl:ex_points}}
\end{table}

\begin{table}[t]
\centering
\scalebox{.85}
{\begin{tabular}{|l|l|l|l|l|l|} 
\hline
 & Hierarchy & \multicolumn{2}{|c|}{Limiting behavior of $\left|\rCh\right|$} &
 Possible realization & LSP constraint \\
& &\multicolumn{1}{|c|}{$I=1$} &  \multicolumn{1}{|c|}{$I=2$} & & \\
\hline
I & $x_{k1}, x_{k2} \gg 1$ & $2\frac{\FVAbs}{m^2_{\Cpm_k}}$ & $\frac{\FVAbs}{m^2_{\Cpm_k}}$ & $m_{\tilde L_{11}}, m_{\tilde L_{22}} \ll M_2, \mu$ & \\
%& & & &&\\
II & $x_{k1}, x_{k2} \ll 1$ & $\frac{\FVAbs}{m^2_{\tilde\nu_1}}$ & $\frac{\FVAbs}{|m^2_{\tilde\nu_2}-m^2_{\tilde\nu_1}\!|} \left| 1-\frac{x_{k1} \log x_{k1}}{x_{k2} \log x_{k2}} \right|$ & $\mu \ll m_{\tilde L_{11}}, m_{\tilde L_{22}}$ & $\mu < m_{\tilde L_{11}}$ \\
III & $x_{k1} \ll 1 \ll x_{k2}$ & $\frac{\FVAbs}{m^2_{\tilde\nu_1}}$ & $\frac{\FVAbs}{ m^2_{\tilde\nu_1}}$ & $ m_{\tilde L_{11}} \gg M_2,\mu \gg  m_{\tilde L_{22}}$ & $M_1< m_{\tilde L_{22}}$ \\
IV & $x_{k2} \ll 1 \ll x_{k1}$ & $\frac{1}{2} \frac{\FVAbs}{m^2_{\Cpm_k}}$  & $\frac{1}{2} \frac{\FVAbs}{m^2_{\tilde\nu_2}} \left| 2 + \frac{1}{x_{k2} \log x_{k2}} \right|$ & $m_{\tilde L_{11}} \ll \mu \ll  m_{\tilde L_{22}}$  & \\ 
V & $x_{k1} \sim x_{k2} \sim 1$ & $\frac{2}{5} \frac{\FVAbs}{m^2_{\tilde\nu_1}}$ & $\frac{1}{4} \frac{\FVAbs}{m_{\tilde \nu_2}^2} $    & $ \mu \sim m_{\tilde L_{11}} \sim m_{\tilde L_{22}}$  & $M_1 < m_{\tilde L_{11}}, m_{\tilde L_{22}}$ \\
VI & $x_{k1} \sim 1$, $x_{k2} \gg 1$ & $\frac{1}{2}\frac{\FVAbs}{m^2_{\tilde\nu_1}}$ & $\frac{1}{3} \frac{\FVAbs}{m_{\tilde \nu_1}^2} $   & $\mu \sim m_{\tilde L_{11}} \gg m_{\tilde L_{22}} $  &    \\
VII & $x_{k1} \sim 1$, $x_{k2} \ll 1$ & $\frac{1}{4}\frac{\FVAbs}{m^2_{\Cpm_k}}$ & $\frac{1}{3}\frac{\FVAbs}{m^2_{\Cpm_k}} \,/ \log x_{k2}^{-1}$ & $\mu \sim m_{\tilde L_{11}} \ll m_{\tilde L_{22}}$ & $\mu < m_{\tilde L_{11}}$ \\ 
VIII & $x_{k2} \sim 1$, $x_{k1} \gg 1$ & $\frac{\FVAbs}{m_{\tilde \nu_2}^2}$ 
& $\frac{1}{2}\frac{\FVAbs}{m^2_{\tilde\nu_2}}$ 
    & $m_{\tilde L_{11}} \ll \mu \sim m_{\tilde L_{22}}$ & 
     \\
IX & $x_{k2}\sim 1$, $x_{k1} \ll 1$ & $\frac{\FVAbs}{m^2_{\tilde\nu_1}}$ & $\frac{\FVAbs}{m^2_{\tilde\nu_1}}$ & $\mu \sim m_{\tilde L_{22}} \ll m_{\tilde L_{11}}$ & $M_2$ or $\mu < m_{\tilde L_{22}}$ \\
\hline
\end{tabular}
}
\caption{Different limits of the ratio $\rChS$, according to different
hierarchies of the parameters $x_{ki}$, for $I=1,2$. 
These approximations are valid chargino by chargino, thus one may have
to consider different hierarchies for the two charginos. The last
column shows the constraints to obtain a neutralino LSP; where no
constraint is given, this is not possible.
\label{tbl:ch_correlations} }
\end{table}

\subsection{Chargino dominance}
\label{sec:chargino dominance}
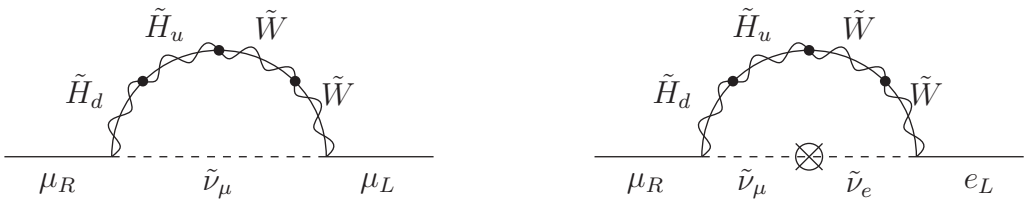
\begin{figure}
\centering
\begin{picture}(480,100)(0,0)
\Line(40,30)(80,30)
\CArc(120,30)(40,360,180)
\PhotonArc(120,30)(40,0,180){3}{8}
\Vertex(120,70){2}
\Vertex(91.6,58.4){2}
\Vertex(148.4,58.4){2}
\DashLine(80,30)(160,30){3}
\Line(160,30)(200,30)
\Text(60,20)[1]{$\mu_R$}
\Text(120,20)[1]{$\tilde\nu_\mu$}
\Text(180,20)[1]{$\mu_L$}
\Text(100,80)[1]{$\tilde H_u$}
\Text(140,80)[1]{$\tilde W$}
\Text(70,55)[1]{$\tilde H_d$}
\Text(165,55)[1]{$\tilde W$}
\Line(260,30)(300,30)
\CArc(340,30)(40,360,180)
\PhotonArc(340,30)(40,0,180){3}{8}
\Text(320,80)[1]{$\tilde H_u$}
\Text(360,80)[1]{$\tilde W$}
\Text(290,55)[1]{$\tilde H_d$}
\Text(385,55)[1]{$\tilde W$}
\Vertex(340,70){2}
\Vertex(311.6,58.4){2}
\Vertex(368.4,58.4){2}
\DashLine(300,30)(380,30){3}
\Line(380,30)(420,30)
\Line(335,35)(345,25)
\Line(335,25)(345,35)
\Text(280,20)[1]{$\mu_R$}
\Text(320,20)[1]{$\tilde \nu_\mu$}
\Text(360,20)[1]{$\tilde \nu_e$}
\Text(405,20)[1]{$e_L$}
\CArc(340,30)(5,0,360)
\end{picture}
\caption{Mass-insertion diagrams involving higgsinos $\tilde{H}_{u,d}$
  and winos $\tilde{W}$, which correspond to the leading contributions to $a_\mu$  and
  $\amuegG$ in the case of chargino dominance. The external photon can
couple to all charged lines. The cross denotes the insertion of the
flavor mixing term $m_{\tilde L_{12}}^2$.}
\label{eq:spCh_diagrams}
\end{figure}

We first focus on  cases in which the chargino contributions are dominant.
For large $\tan\beta$ they
are essentially given by $a_\mu^{\tilde\chi^\pm_k}{}_2$
and $a^{\tilde \chi^{\pm}_k}_{\mu e \gamma\ 2}$;  the corresponding
mass-insertion diagrams are shown in \Figref{eq:spCh_diagrams}.
The chargino contributions depend only on four free mass parameters, $\mu$,
$M_2$, $m_{\tilde L_{11}}$, $m_{\tilde L_{22}}$, and the mixing
$m_{\tilde L_{12}}$. 

The correlations are governed by \eq{eq:corr_mueg_mu2}, which
relates the contributions to $a_\mu$ and $\amuegL$ from one individual
chargino $\tilde{\chi}^\pm_k$ and has two immediate implications. The
ratio $\rChS$ in \eq{eq:corr_mueg_mu2} depends non-trivially on the mass ratios $x_{k1}$ and 
$x_{k2}$; hence (i) it is different for the two charginos and the
correlation between the sums over the contributions can be much weaker
than the correlations between the individual contributions, and
(ii) even for the individual charginos the right-hand side depends on the
mass hierarchy between charginos and sneutrinos.

Table \ref{tbl:ch_correlations} shows a comprehensive list of hierarchies
between chargino and sneutrino masses, following Ref.\ \cite{Chacko:2001xd}. For each
hierarchy we analytically evaluated the limiting behavior of the
ratio $\rChS$. The results are
shown in the third and fourth column of the table; they slightly
improve similar 
results of Ref.\ \cite{Chacko:2001xd}. Note that the result for $I=2$
is the more relevant one since the $I=2$ contributions dominate for
large $\tan\beta$.
The last column shows the conditions that are necessary for having a
neutralino LSP\@.
Note that the lightest neutralino need not be the LSP, if a super-weakly
interacting particle such as the gravitino or axino is the LSP, or if
$R$ parity is violated.
Finally, the table also shows mass patterns which realize the hierarchies and which
allow dominance of the contributions of the lightest or both charginos over the neutralino
contributions. 
In the following we will study each mass hierarchy,
focussing in particular on the deviations from the approximations in
the table and on the 
impact of possible cancellations between individual contributions.

\begin{figure}%[htp]
\centering
\includegraphics{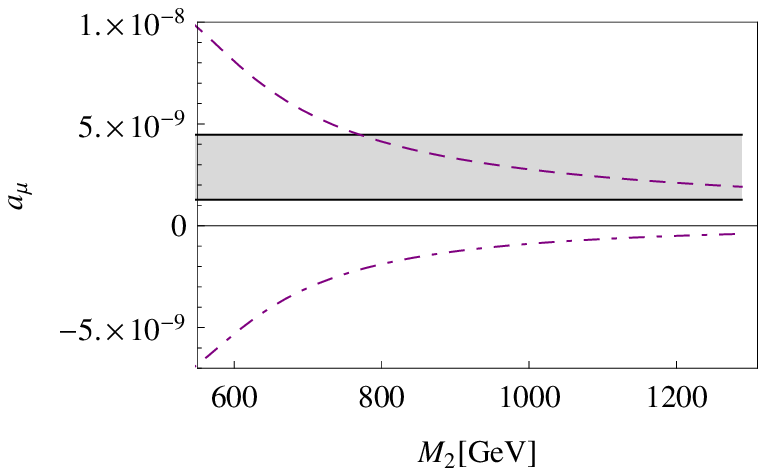}
\includegraphics{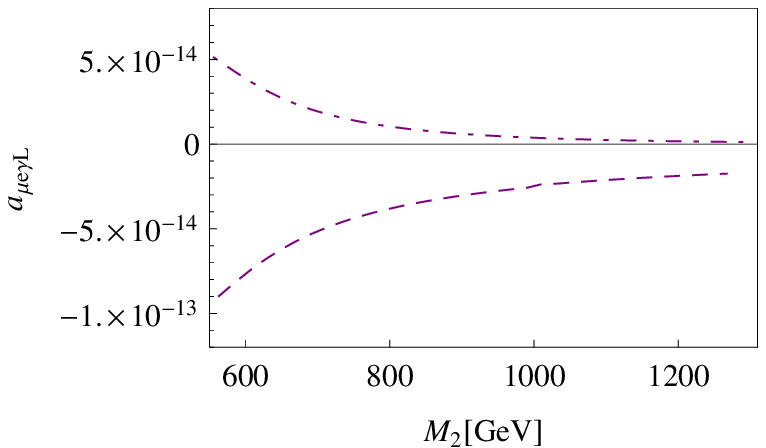}\\[3mm]
\includegraphics{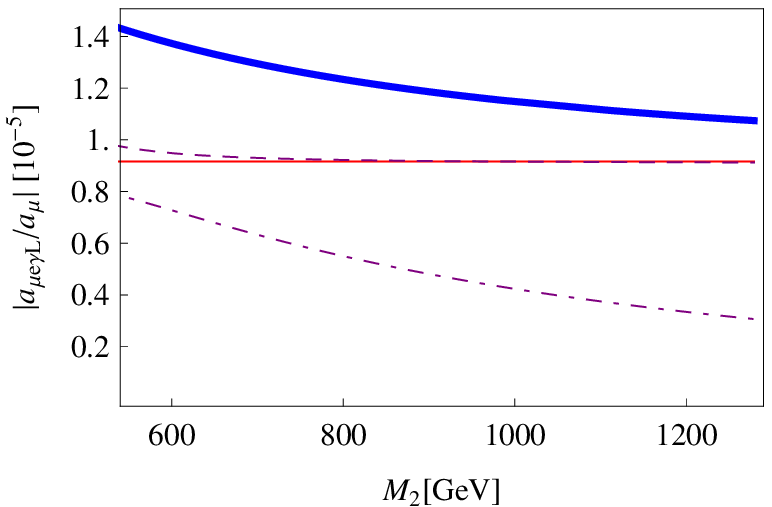}
\caption{$a_\mu$, $\amuegL$ and
their ratio as a function of $M_2$ for the parameters in column II of \Tabref{tbl:ex_points},
equivalently case V in \eq{eq:more_cases_chardom} and \Tabref{tbl:ch_correlations}.
Top: Contributions to $a_\mu$ and $\amuegL$ from each individual chargino.
The long-dashed (dot-dashed) curve corresponds to the lightest (heaviest) chargino.
The shaded area marks the favored $2\sigma$ region
according to \eq{eq:Dmuallw_exp}. Bottom: the ratio $|\amuegL^{\Cpm_k}/a_\mu^{\Cpm_k}|$ for each
chargino (same line styles as in the top row).
For comparison, the corresponding approximation for $|\rChSII|$, case
V in \Tabref{tbl:ch_correlations}, is shown as a thin solid red
line. The total ratio including neutralino contributions, $|\ramTLS|$, is shown as the thick solid blue line.
}
\label{fig:EXIrats}
\end{figure}

Let us consider first the situation of very similar masses, i.e.\ case
V in \Tabref{tbl:ch_correlations},
\begin{equation}
m_{\tilde L_{11}} \sim m_{\tilde L_{22}} \sim \mu 
\end{equation}
or
\begin{equation}
x_{11}\sim x_{12} \sim 1, 
\end{equation}
in terms of the mass ratios $x_{ki}$ defined in \eq{eq:defratioXki}.
As a concrete example, we choose the parameters appearing in 
column II of \Tabref{tbl:ex_points}. The chargino masses are driven by $\mu$ and $M_2$, respectively,
$m_{\Cpm_1}\approx 500\GeV$ and $m_{\Cpm_2} \in [550, 1300] \GeV$. The first and second plot of \Figref{fig:EXIrats} show, respectively, the
contributions to $a_\mu$ and $\amuegL$ from each chargino and demonstrate that the
lightest chargino provides the dominant contributions. The larger
$M_2$, the more pronounced the domination becomes. The third plot of
the same figure shows the ratios of the contributions from each
chargino individually and the ratio of the sums of all contributions
to $a_\mu$ and $\amuegL$. The ratios are compared with the prediction
for the theoretical
limiting behavior given in \Tabref{tbl:ch_correlations},
$|\ramTL| = \frac{1}{4} \frac{\FVAbs}{m^2_{\tilde\nu_2}}$. Indeed, the
ratio for the lightest chargino agrees excellently with this
prediction, because the hierarchy condition
$x_{11} \sim x_{12} \sim 1$ is satisfied precisely. For the heaviest
chargino, $x_{21}$ and $x_{22}$ are significantly larger than $1$, except
at the lower end of the considered interval for $M_2$, so for higher $M_2$ the corresponding ratio
deviates more and more from
$\frac{1}{4} \frac{\FVAbs}{m^2_{\tilde\nu_2}}$.
Due to the dominance of the lightest chargino, the ratio
of the sum of all contributions agrees with the
prediction within a factor $1.5$ for all values of $M_2$ shown in the plot.
The absolute value of this ratio, $|\ramTLS|$, is larger than both
individual ratios $|\amuegL^{\Cpm_k}/a_\mu^{\Cpm_k}|$, which indicates
that the cancellation between the different contributions is
stronger for $a_\mu$ than for $\amuegL$.
 As $M_1$ is the smallest of the soft SUSY breaking parameters, the
LSP is a neutralino with dominant bino component.

We will now study systematically all nine cases listed in \Tabref{tbl:ch_correlations}. We
set the mass parameters to definite values according to the
hierarchies given in the 
table, but such that the lightest chargino still represents the
leading contribution to $a_{\mu}$, and such that the
value of $a_\mu$ remains inside the region allowed by
\eq{eq:Dmuallw_exp}.
We use the following sets of parameters:
\begin{equation}
 \begin{array}{lllll}
 {\rm{I.}}  &  m_{\tilde \nu_e}= 127 \ \rm{GeV}, & m_{\tilde\nu_\mu}= 117 \ \rm{GeV}, & \mu=550\  {\rm{GeV}},&  M_1=700\ \rm{GeV},\\
  {\rm{II.}}  &  m_{\tilde \nu_e}= 489 \ \rm{GeV}, & m_{\tilde\nu_\mu}= 693 \ \rm{GeV}, &\mu=220\  {\rm{GeV}}, &  M_1=700\ \rm{GeV},\\
  {\rm{III.}} &   m_{\tilde \nu_e}= 3500 \ \rm{GeV}, & m_{\tilde\nu_\mu}= 356 \ \rm{GeV}, &\mu=1400\  {\rm{GeV}}, & M_1=350 \ {\rm{GeV}}, \\
{\rm{IV.}} &   m_{\tilde \nu_e}= 96\ \rm{GeV}, & m_{\tilde\nu_\mu}= 965\ \rm{GeV}, & \mu=320\  {\rm{GeV}}, & M_1=400 \ {\rm{GeV}}, \\
\rm{V.} &   m_{\tilde \nu_e}= 496\  \rm{GeV}, & m_{\tilde\nu_\mu}= 496\  \rm{GeV}, & \mu=500\  {\rm{GeV}}, & M_1=400 \ \rm{GeV}, \\
 {\rm{VI.}} &  m_{\tilde \nu_e}= 797 \ \rm{GeV}, & m_{\tilde\nu_\mu}= 131 \ \rm{GeV}, & \mu=800\  {\rm{GeV}}, &  M_1=700 \   {\rm{GeV}},\\
 {\rm{VII.}} &   m_{\tilde \nu_e}= 256 \ \rm{GeV}, & m_{\tilde\nu_\mu}= 917 \ \rm{GeV} &\mu=260\  {\rm{GeV}}, & M_1=500 \ {\rm{GeV}},\\
{\rm{VIII.}} &   m_{\tilde \nu_e}=66 \ \rm{GeV}, & m_{\tilde\nu_\mu}= 550 \ \rm{GeV} &\mu=550 \  {\rm{GeV}},   & M_1=500 \ {\rm{GeV}}, \\
 {\rm{IX.}} &   m_{\tilde \nu_e}= 1738 \ \rm{GeV}, & m_{\tilde\nu_\mu}= 520 \ \rm{GeV}, &  \mu=500 \ {\rm{GeV}}, &M_1=900 \   {\rm{GeV}}. 
\end{array}
\label{eq:more_cases_chardom}
\end{equation}
Here we have used $m_{\tilde \nu_e}^2 \equiv m^2_{\tilde L_{11}} +
\mathcal{D}^\nu_L$ and $m_{\tilde\nu_\mu}^2 \equiv m^2_{\tilde L_{22}} +
\mathcal{D}^\nu_L$ as more physical inputs.
In addition, $\tan\beta=50$, $m^2_{\tilde L_{12}} = (3\GeV)^2$,
and the value of $M_2$ is kept as a variable.

\begin{figure}[t]
\centering
\includegraphics{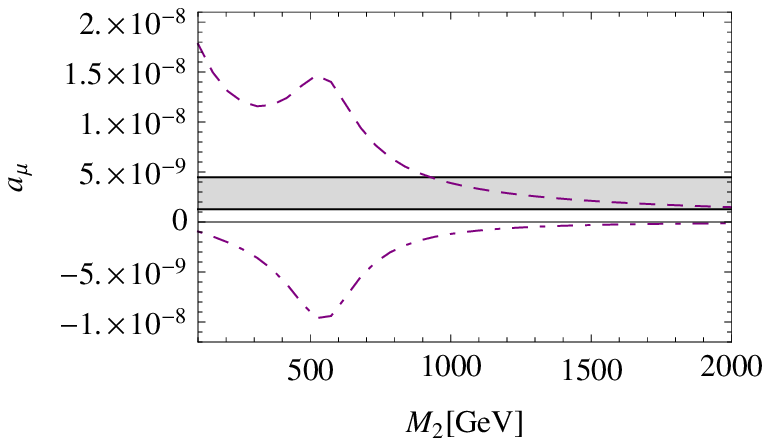}
\includegraphics{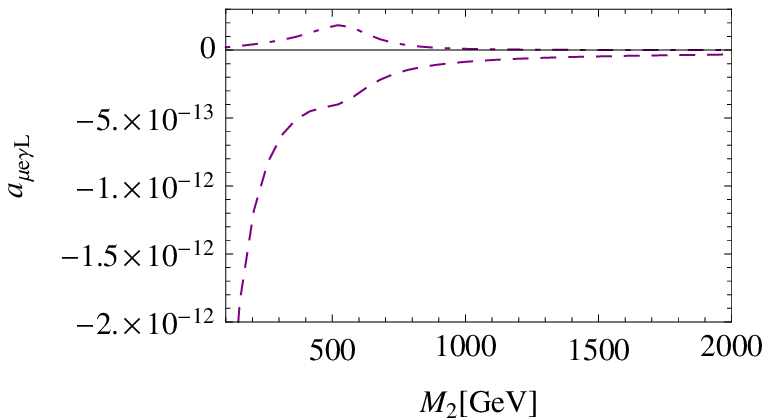}\\[3mm]
\includegraphics{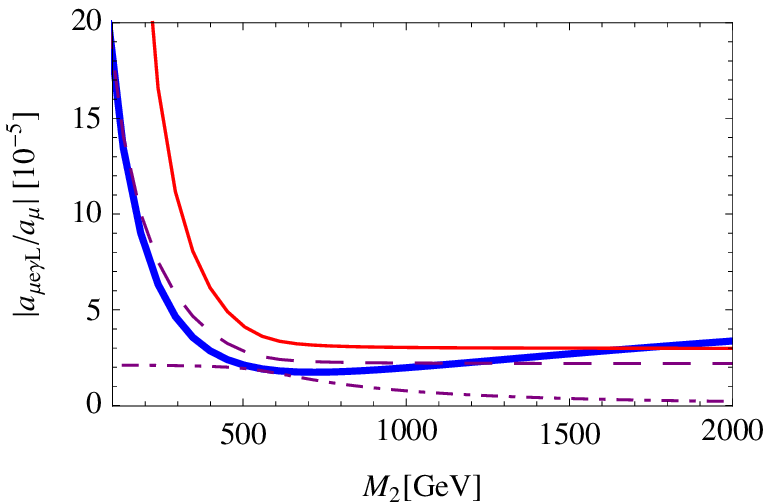}
\caption{Case I\@.  Same as in \Figref{fig:EXIrats}, but for case I in
  \Tabref{tbl:ch_correlations} and \eq{eq:more_cases_chardom}.
  The thin solid red line corresponds to the approximation for the
  lightest chargino, $k=1$.
\label{fig:ch_correlations1}}
\end{figure}

For each case, we focus on the following questions: Does the limiting
behavior in \Tabref{tbl:ch_correlations}  
provide a reliable prediction for
the correlation between $a_\mu$ and $\amuegG$ for (i) the individual
contribution of the lighter chargino and
(ii) the sum of all contributions? We consider the prediction reliable if it agrees with the
precise numerical result up to a factor of $1.5$ or less.
In addition, (iii) are the scenarios
compatible with a neutralino LSP\@? We will be briefer in our
discussions than for case~V discussed above, but we will highlight cases I,
VI, and VII, each of which illustrates a different behavior. 
Case V is an example where all three criteria are met. In case VI, (i) and (ii) are satisfied but (iii) is not.
Case I is an example where (i) is satisfied whereas (ii) 
is met only in a part of the considered mass range for $M_2$.
Case VII is an example where only (i) and (iii) are satisfied; in
addition, the approximation to
$\amuegII/\amuII$ 
cannot be written as a ratio of two parameters, because there are logarithmic contributions that become relevant. For these cases we also show plots
analogous to \Figref{fig:EXIrats}, see Figs.~\ref{fig:ch_correlations1}--\ref{fig:ch_correlations7}.

\begin{itemize}
\item
For case I, both $\mu$ and $M_2$ must be considerably bigger than the
sneutrino masses to generate the hierarchy $x_{k1}, x_{k2} \gg 1$, which
pushes the masses of both charginos and three neutralinos well above the
masses of the sneutrinos. In order to guarantee chargino dominance, we
also set $M_1$ to a high value, which implies that in case I we cannot
obtain a neutralino LSP\@.
\Figref{fig:ch_correlations1} shows that the approximation
for the ratio $\amuegII/\amuII$ works within $50\%$ accuracy for
the lightest chargino.  It fails in the low-$M_2$ region
where $x_{1i}$ becomes smaller than about $8$, 
since there the expansion of the ratio in \eq{eq:corr_mueg_mu} is not
valid anymore.  
As it depends on the chargino mass, the approximation for $k=1$
plotted in the figure cannot describe the contribution of the heavier
chargino.
Besides, at $\mu=M_2$ the chargino mixing becomes maximal, and
so does the cancellation between the chargino contributions. Hence,
approximating one of them well can be meaningless for the sum. 
The approximation works reliably for the sum only for $M_2 \gtrsim 1\TeV$.

\item
For case II, the approximation in \Tabref{tbl:ch_correlations} contains a
factor involving logarithms. This factor can change the result by more than a factor of 2
% if the sneutrino masses are not similar,
and therefore should not be neglected. If it is included, the approximation works very well for the
contribution of the lighter chargino and satisfactorily for the sum.
%For this case the identity of the LSP is not subject to conditions on
%$M_2$, so $M_2$ can be even bigger than $1\TeV$.
For the specific example in \eq{eq:more_cases_chardom} and $M_2\gg \mu$,
the LSP is a higgsino-like neutralino with a mass around $216 \GeV$. 

\item
For case III the approximation works quite well for the sum and for
the contribution of the lighter chargino for all values of
$M_2$ up to about $1\TeV$. It is no problem to consider relatively small values of
$M_1$, so the LSP can be a bino-like neutralino.
For the values quoted in \eq{eq:more_cases_chardom}, the LSP has a mass
around $350\GeV$.

\item
Case IV interchanges the roles of $\snu_e$ and $\snu_\mu$
compared to case III\@. Due to the higher
    $m_{\tilde\nu_\mu}$ the contribution to both observables is
    suppressed,  and the
    approximation for the ratio $\amuegL/a_\mu$  depends in a
complicated way on the chargino and $\snu_\mu$ 
masses. We find that even the full approximation shown in
\Tabref{tbl:ch_correlations} does not work reliably.
The hierarchy required in case IV forbids very light values
of $\mu$ and $M_2$ around $100 \GeV$, and the requirement of chargino dominance
forbids values of $M_1$ smaller than around $400\GeV$. This case thus does not allow a neutralino LSP.

\item
For case V, see the discussion above.
\item
For case VI, both $\mu$ and  $M_2$ must
be kept large to generate  the hierarchies $x_{k2}\gg 1$,
$x_{k1}\sim 1$.   If this is satisfied, the approximation works
well for the lighter chargino.  It also provides a reliable prediction for the sum of the contributions with about $40\%$ accuracy
for all values $M_2>800\GeV$, see \Figref{fig:ch_correlations6}. The bino mass should be heavy,
$M_1 \gtrsim 500\GeV$, to avoid neutralino contributions to be dominant, so
with the hierarchy of this case it is not possible to obtain a
neutralino LSP.   

\begin{figure}[t]
\centering
\includegraphics{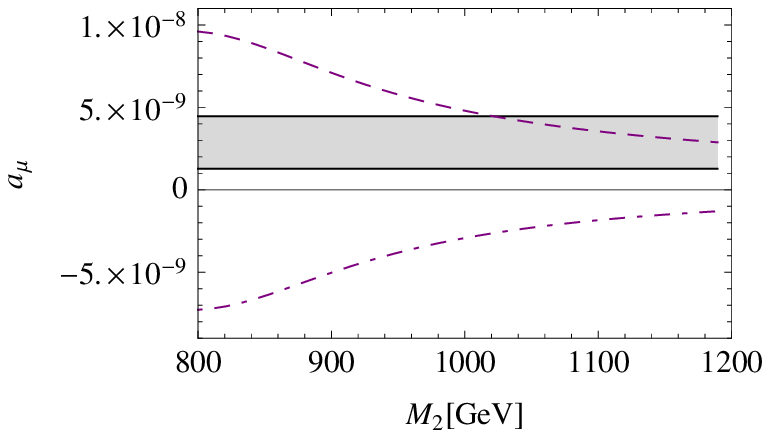}
\includegraphics{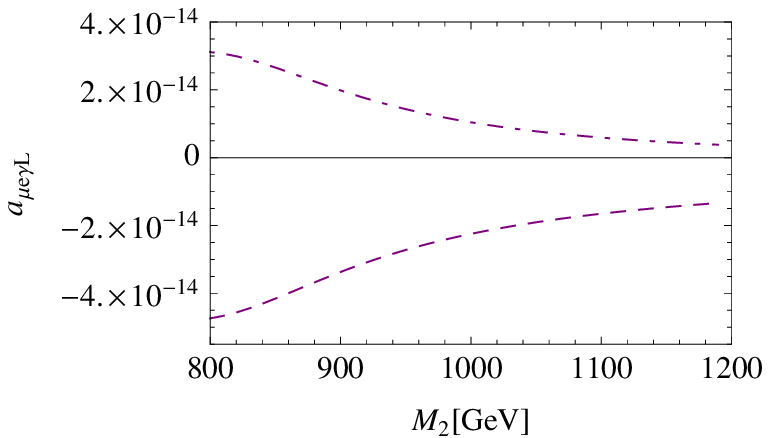}\\[3mm]
\includegraphics{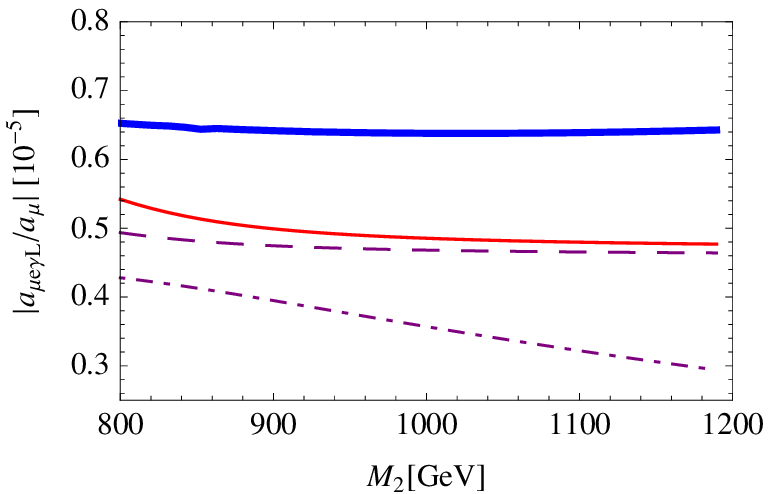}
\caption{Case VI\@. Same as in \Figref{fig:EXIrats}, but for case VI in
\Tabref{tbl:ch_correlations} and \eq{eq:more_cases_chardom}.
\label{fig:ch_correlations6}}
\end{figure}

\item
Case VII is shown in \Figref{fig:ch_correlations7}.   The
plotted range for $M_2$ is the one where  $a_\mu$ is within the
favored $2\sigma$ region.  The $\mu$ parameter is small enough to
satisfy the LSP condition $m_{\tilde\chi^0_1} < m_{\snu_1}$. 
In this case it is important not to neglect the logarithmic factor in
the approximation for $\amuegII/\amuII$, 
which then works very well for the lightest chargino.  
 However, interestingly the sum of the contributions is not sufficiently dominated
by the lightest chargino, so the approximation fails to predict the
ratio $\ramTLS$ correctly.

\begin{figure}[t]
\centering
\includegraphics{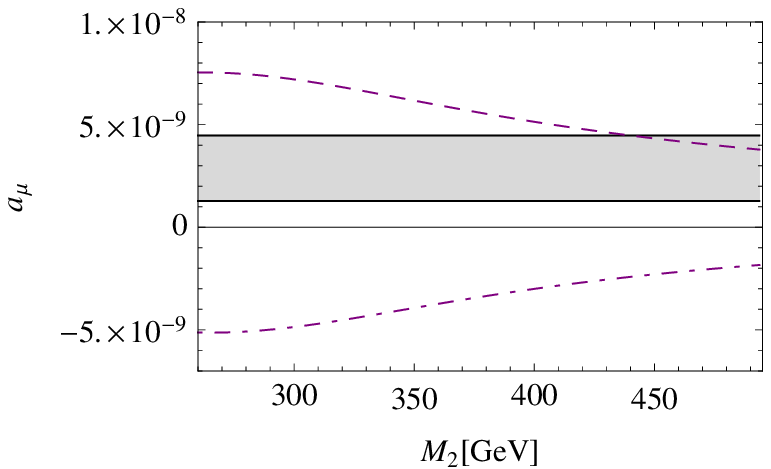}
\includegraphics{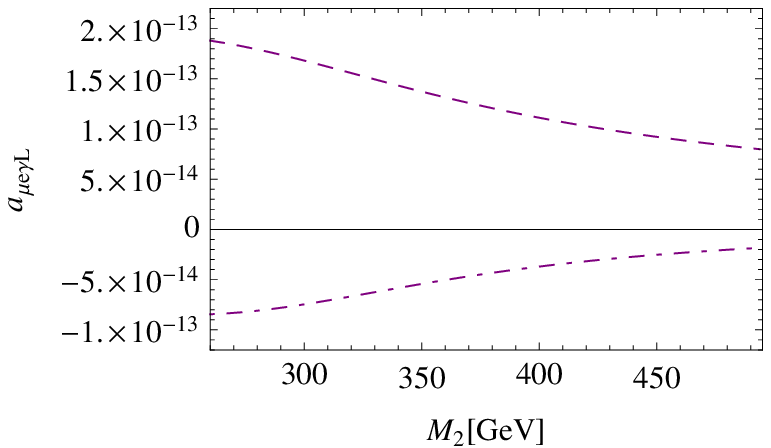}\\[3mm]
\includegraphics{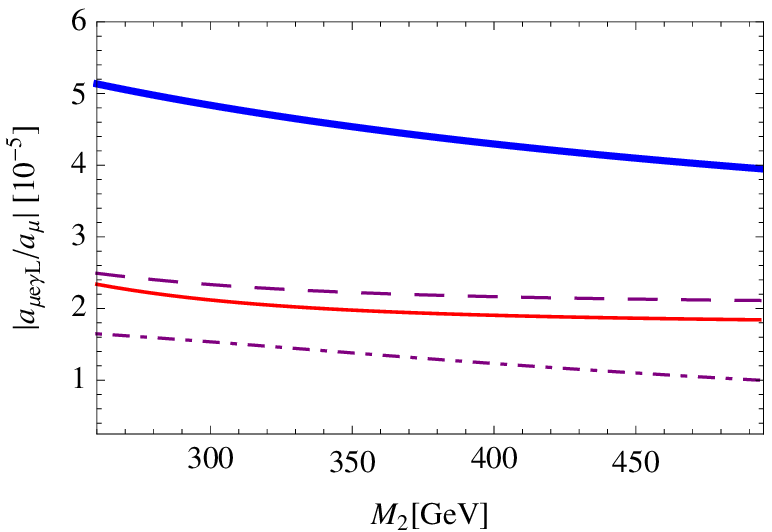}
\caption{Case VII\@. The same as in \Figref{fig:EXIrats}, but for case
  VII  in \Tabref{tbl:ch_correlations} and
  \eq{eq:more_cases_chardom}.
\label{fig:ch_correlations7}}
\end{figure}

\item
Case VIII is similar to case VI in that both $\mu$ and $M_2$ must be kept large to generate the
hierarchies, 
 which together with the requirement of subdominant neutralino
contributions makes it impossible to obtain a neutralino LSP\@.
The approximation for $\amuegII/\amuII$ 
works well for the contribution of the lightest chargino 
 but not for the sum, which shows a strong dependence on $M_2$.

\item
For case IX, the approximation of the ninth line in
\Tabref{tbl:ch_correlations} works quite well for both the lightest
chargino contribution and for the sum of the contributions from the two charginos.
In this case it is possible to obtain a neutralino LSP\@.
For the example of \eq{eq:more_cases_chardom}, it could be higgsino- or wino-like,
depending on the value of $M_2$ in comparison to $\mu$, which remains
fixed at $500\GeV$.
\end{itemize}

Note that for all these cases the values of $m_{\tilde R_{11}}$
and $m_{\tilde R_{22}}$ are not important. These parameters just have to
be sufficently large to avoid the appearance of a charged LSP.

In summary, depending on the hierarchy of the sneutrino and
chargino masses, the different approximations to $\amuegII/\amuII$
given in \Tabref{tbl:ch_correlations}
predict the ratio of the lighter chargino's contributions to $\amuegL$
and $a_\mu$ reliably, i.e.\ within a factor $1.5$ of accuracy,
with case IV being the sole exception.
In cases I (for $M_2 \gtrsim 1\TeV$), II, III, V, VI, and IX, these
approximations can also
be used as a reliable substitute for the exact
value of $\amuegL/a_\mu$ stemming from all contributions.
 However, in case II the approximation depends on three
superparticle masses, weakening the link between the two observables.
In the remaining cases I, III, V, VI, and IX, there is a strong
correlation; we can predict $\BR$ as a function of $a_\mu$ and the ratio
of the flavor-violating parameter and a single superpartner mass within
a factor of roughly $1.5^2$.

\begin{figure}[t]
\centering
\includegraphics[width=0.45\textwidth]{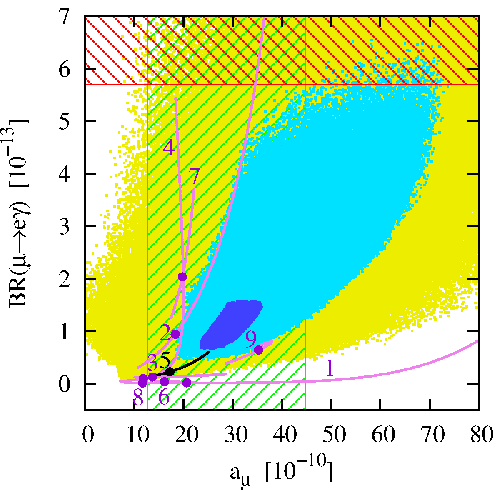}
\caption{Comparison of our chargino dominance benchmark points to the general scatter plots of $\BR$ versus $a_\mu$.
The numbered dots correspond to the benchmark points listed in \eq{eq:more_cases_chardom},
with $M_2$ chosen as the midpoint of the ranges shown in
Figs.~\ref{fig:EXIrats}--\ref{fig:ch_correlations7}.  The curves
arise from the variation of $M_2$ within these ranges.
For cases II--IV, $M_2\in[100,1500]\GeV$, for case~VIII,
$M_2\in[550,2050]\GeV$, and for case IX, $M_2\in[80,400]\GeV$.
The hatched vertical band depicts the $2\sigma$ range of $\Delta a_\mu$,
and the hatched top region is excluded at the 90$\%$ C.L.\ by MEG\@.
For the general scatter plots we have fixed $\deltaLL = \deltaRR = 2 \times 10^{-5}$ and selected SUSY masses in the regions $[430,530]\GeV$ (dark blue), $[300,600]\GeV$ (light blue) and $[200,1000]\GeV$ (yellow).}
\label{fig:ScatterEqualM}
\end{figure}

Figure \ref{fig:ScatterEqualM} summarizes the present section and allows
further interpretation of the similar-mass case of the previous section
\ref{sec:SimilarMasses}. It shows $\BR$ versus $a_\mu$ for the
benchmark points listed in \eq{eq:more_cases_chardom} and for random
SUSY masses. The scatter regions repeat the similar-mass case of
\Figref{fig:ScatterEqualM-R}, but this time focussing on the effect
of different mass intervals instead of the strength of cancellation.
The random SUSY masses are generated in the regions 
$[430,530]\GeV$ (dark blue region), $[300,600]\GeV$ (light blue) and
$[200,1000]\GeV$ (yellow). We have fixed $\deltaLL = \deltaRR = 2
\times 10^{-5}$.

The benchmark points represent parameter choices with certain extreme
mass hierarchies, each of which leads to a different correlation
between $\BR$ and $a_\mu$, as we have discussed in this section. Indeed, one finds a wide variety of
branching ratios even for almost 
the same values of $a_\mu$,
see e.g.\ dots 2, 5, and 6, which are nearly aligned on
a common vertical line.

Given those drastically different limits of the ratio,
the reason becomes transparent for the wide spread
of the points in the scatter regions, where the masses are similar but
not exactly equal.

\subsection[Large mu limit]{Large $\mu$ limit} \label{sec:LargeMu}

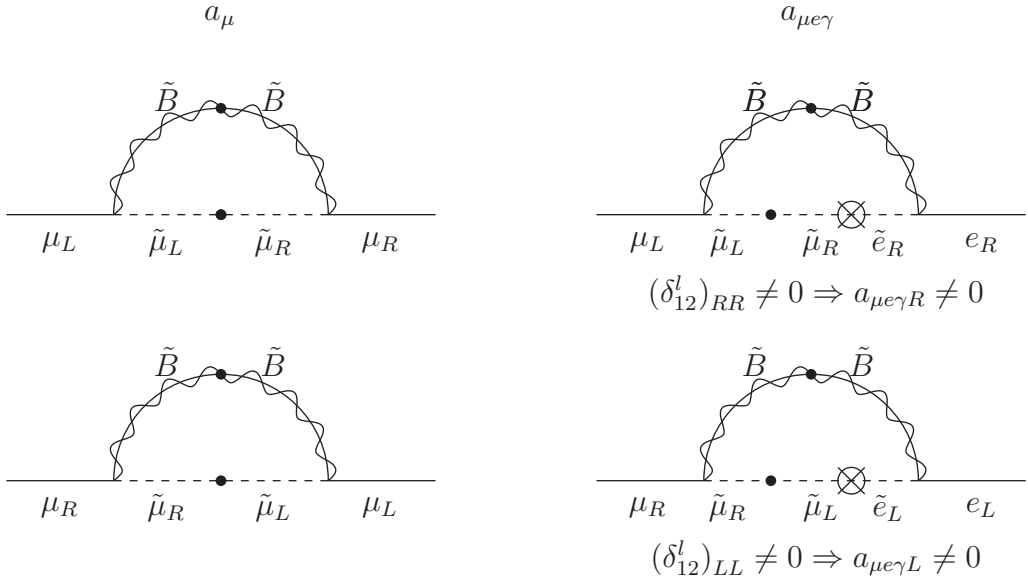
\begin{figure}
\centering
\begin{picture}(480,215)(0,0)
\Text(120,205)[1]{$a_\mu$}
\Line(40,130)(80,130)
\CArc(120,130)(40,360,180)
\PhotonArc(120,130)(40,0,180){3}{8}
\Vertex(120,170){2}
%\Line(335,75)(345,65)
%\Line(335,65)(345,75)
%
\Vertex(120,130){2}
%\DashArrowLine(80,130)(160,130){3}
\DashLine(80,130)(160,130){3}
\Line(160,130)(200,130)
\Text(60,120)[1]{$\mu_L$}
\Text(100,120)[1]{$\tilde\mu_L$}
\Vertex(120,130){2}
\Text(140,120)[1]{$\tilde\mu_R$}
\Text(180,120)[1]{$\mu_R$}
%\Line(115,175)(125,165)
%\Line(115,165)(125,175)
\Text(100,175)[1]{$\tilde B$}
\Text(140,175)[1]{$\tilde B$}
\Text(340,205)[1]{$ a_{\mu e \gamma}$}
\Vertex(340,170){2}
%\Line(335,175)(345,165)
%\Line(335,165)(345,175)
%
\Line(260,130)(300,130)
\Text(320,175)[1]{$\tilde B$}
\Text(360,175)[1]{$\tilde B$}
%\DashArrowLine(300,130)(380,130){3}
\DashLine(300,130)(380,130){3}
\CArc(340,130)(40,360,180)
\PhotonArc(340,130)(40,0,180){3}{8}
\Line(380,130)(420,130)
\Vertex(325,130){2}
\Line(350,135)(360,125)
\Line(350,125)(360,135)
\Text(280,120)[1]{$\mu_L$}
\Text(310,120)[1]{$\tilde\mu_L$}
\Text(345,120)[1]{$\tilde\mu_R$}
\Text(370,120)[1]{$\tilde e_R$}
\Text(405,120)[1]{$e_R$}
\Text(343,101)[1]{$\deltaRR \neq 0 \Rightarrow \amuegR \neq 0$}
\Line(40,30)(80,30)
\CArc(120,30)(40,360,180)
\PhotonArc(120,30)(40,0,180){3}{8}
%\Vertex(120,70){2}
\Text(320,175)[1]{$\tilde B$}
\Text(360,175)[1]{$\tilde B$}
%
%\DashArrowLine(80,30)(160,30){3}
\DashLine(80,30)(160,30){3}
\Line(160,30)(200,30)
\Text(60,20)[1]{$\mu_R$}
\Text(100,20)[1]{$\tilde\mu_R$}
\Vertex(120,30){2}
\Vertex(120,70){2}
\Text(140,20)[1]{$\tilde\mu_L$}
\Text(180,20)[1]{$\mu_L$}
\Text(100,75)[1]{$\tilde B$}
\Text(140,75)[1]{$\tilde B$}
\Line(260,30)(300,30)
\CArc(340,30)(40,360,180)
\PhotonArc(340,30)(40,0,180){3}{8}
\Text(320,75)[1]{$\tilde B$}
\Text(360,75)[1]{$\tilde B$}
%\DashArrowLine(300,30)(380,30){3}
\DashLine(300,30)(380,30){3}
\Line(380,30)(420,30)
\Vertex(325,30){2}
\Vertex(340,70){2}
\Line(350,35)(360,25)
\Line(350,25)(360,35)
\CArc(355,130)(5,0,360)
\CArc(355,30)(5,0,360)
\Text(280,20)[1]{$\mu_R$}
\Text(310,20)[1]{$\tilde\mu_R$}
\Text(345,20)[1]{$\tilde\mu_L$}
\Text(370,20)[1]{$\tilde e_L$}
\Text(405,20)[1]{$e_L$}
\Text(343,1)[1]{$\deltaLL \neq 0 \Rightarrow \amuegL \neq 0$}
\end{picture}
\caption{Leading contributions to $a_\mu$  and $\amuegG$ in the case of large $\mu$.}
\label{eq:spNeut_diagrams}
\end{figure}

If $\tan\beta$ is large, an interesting neutralino contribution to $a_\mu$
and $\amuegG$ is represented by the mass-insertion diagrams in
\Figref{eq:spNeut_diagrams} with bino exchange. These diagrams grow linearly
with $\mu$.  Their contribution to $a_\mu$ is proportional to
$m_\mu^2 \, \mu \tan\beta \, M_1 F(M_1,m_{\tilde\mu_R}, m_{\tilde \mu_L})$,
where $F$ denotes the loop function involved in each diagram. All
other contributions involve higgsinos and are therefore suppressed for
large $\mu$.
Hence, the diagrams of \Figref{eq:spNeut_diagrams} dominate for
sufficiently large $\mu$. We now analyze this parameter region and the
behavior of $a_\mu$ and  $\amuegG$. Very recently, $a_\mu$ in this
scenario has also been studied in detail in 
Ref.\ \cite{Endo:2013lva}.
The most important parameters of the parameter region with large $\mu$
are
$\mu, M_1, m^2_{\tilde L_{22}}, m^2_{\tilde R_{22}}$ and $\tan\beta$.
For our analysis we choose $\tan\beta=50$ and keep all supersymmetric
mass parameters except $\mu$ and $M_2$ 
between $200$ and $600$ GeV.
 We vary $\mu$, focussing 
on the region where
\begin{equation} \label{eq:hierarchiesI}
\mu > M_2 > M_1 .
\end{equation}
Choosing $M_2$ significantly larger than $M_1$ further increases the
dominance of the contribution of the lightest neutralino for large $\mu$.

Note that for large values of $\mu$ and $\tan\beta$, charge-breaking
minima in the scalar potential endanger the stability of the electroweak
vacuum.  This problem can be alleviated by increasing the stau masses
\cite{Endo:2013lva}, which otherwise have no impact on $a_\mu$ and
$\amuegG$.

\begin{figure}%[htp]
\centering
\includegraphics{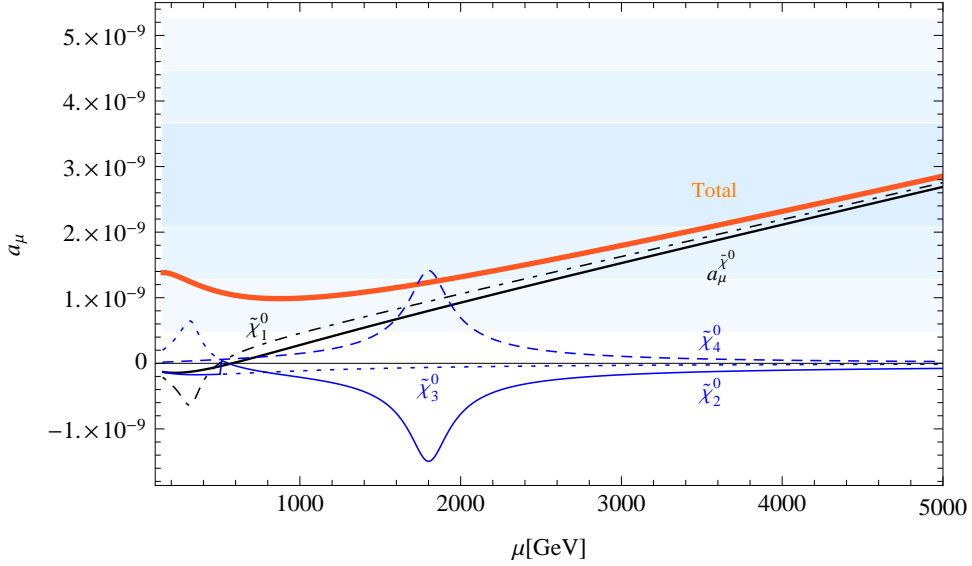}
\caption{{$a_\mu$ as a function of $\mu$ for the large-$\mu$ case of
spectrum III in \Tabref{tbl:ex_points}. The thick solid  (orange)
curve corresponds to the total SUSY contribution. The solid
black line corresponds to the total neutralino contribution, and the
dot-dashed black line to the contribution of the lightest, bino-like
neutralino, which is dominant for large $\mu$. The contributions of the
other neutralinos are shown as {blue lines}.
The horizontal bands represent the experimentally allowed regions at
$1\sigma$, $2\sigma$ and $3\sigma$, respectively.
}
\label{fig:largemuex}
}
\end{figure}

\paragraph{Interplay of contributions to $\boldsymbol{a_\mu}$}
In \Figref{fig:largemuex}, we present an example for this scenario and
show $a_\mu$ as a function of $\mu$ for the parameter choice of spectrum III in
\Tabref{tbl:ex_points}. The thick solid (orange) curve corresponds to
the total SUSY contribution to $a_\mu$. The solid black line
corresponds to the total neutralino contribution and the dot-dashed black
line to the contribution of the lightest (bino-like) neutralino. As we
can see this lightest neutralino almost fully accounts for the total
$a_\mu$ in the large-$\mu$ region, where $\mu$ is significantly larger
than $M_2$. Here the diagrams of \Figref{eq:spNeut_diagrams} dominate.
We have also plotted the other individual neutralino contributions (as
blue solid, dotted, and dashed lines) and
extended the figure down to smaller values of $\mu$, so that it shows the transition from a regime with
chargino dominance at low $\mu$ to the bino-dominated regime.
At very small values of $\mu$, the chargino contributions are
important and the neutralino contributions only
account for a fraction of the total $a_\mu$. At intermediate values, for $\mu \approx M_1$ or
$\mu \approx M_2$, there are two neutralino mass
eigenstates whose contributions {are strongly enhanced due to
large neutralino mixing. 
However, the enhanced contributions roughly cancel each other, so
the sum of all contributions shows no enhancement.  This can be
understood from the mass-insertion diagrams, which have a monotonous
behavior.  Around $\mu\approx500\GeV$, we notice a discontinuity in
the plot, which reflects an exchange of identity of the higgsino-like
mass eigenstates $\tilde\chi^0_2$ and $\tilde\chi^0_3$
that is caused by the mixing between higgsinos and gauginos.
}

In the large $\mu$ limit, the total SUSY contribution can be written as
\begin{equation}
a_\mu \approx g^2_1 \frac{m_\mu}{48 \pi^2} \sum_m \text{Re}[K_{m5}^* K_{m2}] \,\frac{M_1}{m^2_{\tilde \ell_m}} \, F_2^N(x_{1m}) ,
\end{equation}
where $x_{1m}$ is defined in \eq{eq:defratioXim}.
As long as flavor violation is small,
the most important contributions to $a_\mu$ come from the two mass
eigenstates containing mainly $\tilde\mu_L$ and $\tilde\mu_R$, which we
label as $m=b$ and $m=c$.
Then $a_\mu$ can be very well approximated by
\begin{equation}
\label{eq:approxDamMIb}
a_\mu \approx g_1^2 \frac{m_\mu}{48\pi^2}\frac{1}{M_1}
\left\{
\text{Re}[K_{b5}^* K_{b2}] \, x_{1b} F_2^N(x_{1b}) +
\text{Re}[K_{c5}^* K_{c2}] \, x_{1c} F_2^N(x_{1c}) \right\} .
\end{equation}
The factors $K^*_{m5} K_{m2}$ determine the mixing between
$\tilde\mu_L$ and $\tilde\mu_R$, which is a necessary ingredient
as shown by \Figref{eq:spNeut_diagrams}.  As we are dealing with the
mixing of only two states, we can introduce a mixing angle
$\theta_{\tilde\mu}$ with
\begin{equation} \label{eq:ThetaSmu}
K_{b5} K_{b2} \approx -K_{c5} K_{c2} \approx
\frac{1}{2} \sin2\theta_{\tilde\mu} =
\frac{ (m^2_{LR})_{22} }{ m_{\tilde\ell_b}^2-m_{\tilde\ell_c}^2 }
\end{equation}
in analogy to \eq{eq:SneutrinoMixing}, where we have restricted
ourselves to real parameters and chosen the states such that
$m_{\tilde\ell_b} < m_{\tilde\ell_c}$.  We can safely neglect the %muon
trilinear coupling in {the smuon mixing term
$(m^2_{LR})_{22} = m_\mu (A_\mu-\mu^*\tan\beta)$}
{%In the smuon mixing term $(A_\mu-\mu^*\tan\beta)$, see \eq{eq:mass_flav_egs}, we can safely neglect $A_\mu$
and retain only $\mu\tan\beta$.}
Then we can write $a_\mu$ as
\begin{equation} \label{eq:Dmu}
a_\mu \approx
g_1^2 \frac{m_\mu^2}{48\pi^2} \frac{\mu\tan\beta}{M_1} \,
\frac{ x_{1b} F_2^N(x_{1b}) - x_{1c} F_2^N(x_{1c}) }{
 m_{\tilde\ell_c}^2-m_{\tilde\ell_b}^2 } .
\end{equation}
For not too hierarchical smuon masses, the contributions from the two
mass eigenstates are of the same order and partially cancel.  Note that
$x F_2^N(x)$ increases monotonously, so $a_\mu > 0$ for $\mu>0$.
The result for this scenario can be summarized as
\begin{align}
a_\mu \approx {}&
2 \times 10^{-9} \left( \frac{\mu}{4\TeV} \right)
\left( \frac{\tan\beta}{50} \right)
\left( \frac{300\GeV}{M_1} \right)
\left( \frac{1\TeV}{m_{\tilde\ell_c}+m_{\tilde\ell_b}} \right)
\left( \frac{100\GeV}{m_{\tilde\ell_c}-m_{\tilde\ell_b}} \right)
\left( \frac{\Delta(xF^N_2)}{0.1} \right) ,
\nonumber\\
&\Delta(xF^N_2) \equiv x_{1b} F_2^N(x_{1b}) - x_{1c} F_2^N(x_{1c}) .
\end{align}
We could further approximate
\begin{equation}
\frac{\Delta(xF^N_2)}{m^2_{\tilde\ell_c}-m^2_{\tilde\ell_b}} \approx
\frac{M_1^2}{m^2_{\tilde\ell_b}m^2_{\tilde\ell_c}} \,
\frac{d}{dx} x F_2^N(x) \Big|_{x_{1a}} ,
\end{equation}
where the numerical value of the derivative varies slowly around $1$ in
the region of interest.  Thus, increasing $M_1$ increases $a_\mu$ if all
other parameters are kept fixed.

\paragraph{Interplay of contributions to $\boldsymbol{a_{\mu e\gamma}}$} 
We consider small flavor-violating terms $m^2_{\tilde L_{12}}$ and
$m^2_{\tilde R_{12}}$,
\begin{equation} \label{eq:mL12hier}
 \frac{m^2_{\tilde L_{12}}}{m^2_{\tilde L_{22}}} \ll 1 , \quad
 \frac{m^2_{\tilde R_{12}}}{m^2_{\tilde R_{22}}} \ll 1 .
\end{equation}
For large $\mu$, we have then
\begin{eqnarray}
\amuegL &\approx& g_1^2 \frac{m_\mu}{48\pi^2} \sum_m K^*_{m1} K_{m5} \,
\frac{M_1}{m^2_{\tilde \ell_m}} \, F_2^N(x_{1m}) ,
\label{eq:amuegLLargeMu}
\\
\amuegR &\approx& g_1^2 \frac{m_\mu}{48\pi^2} \sum_m K^*_{m4} K_{m2} \,
\frac{M_1}{m^2_{\tilde \ell_m}} \, F_2^N(x_{1m}) .
\end{eqnarray}
Let us first study the case $m^2_{\tilde R_{12}}=0$ and
$m^2_{\tilde L_{12}}\neq 0$, which implies that only $\amuegL$ is
non-negligible.  As we can see from the corresponding diagram in
\Figref{eq:spNeut_diagrams}, $\amuegL$ receives contributions from the
three mass eigenstates containing mainly $\tilde e_L$, $\tilde\mu_L$
and $\tilde\mu_R$.%
\footnote{If there was a large flavor-violating entry in $m^2_{LR}$,
two eigenstates (those containing mainly $\tilde\mu_R$ and $\tilde e_L$)
could dominate.}
Let us denote them by $\tilde\ell_a$, $\tilde\ell_b$
and $\tilde\ell_c$, respectively.
For these states we estimate
\begin{eqnarray}
K_{a1} K_{a5} &\approx&
\theta_{\tilde\mu} \left(
\frac{ m^2_{\tilde L_{12}} }{ m_{\tilde\ell_c}^2-m_{\tilde\ell_a}^2 } -
\frac{ m^2_{\tilde L_{12}} }{ m_{\tilde\ell_b}^2-m_{\tilde\ell_a}^2 }
\right),
\\
K_{b1} K_{b5} &\approx&
\theta_{\tilde\mu}
\frac{ m^2_{\tilde L_{12}} }{ m_{\tilde\ell_b}^2-m_{\tilde\ell_a}^2 } ,
\\
K_{c1} K_{c5} &\approx&
-\theta_{\tilde\mu}
\frac{ m^2_{\tilde L_{12}} }{ m_{\tilde\ell_c}^2-m_{\tilde\ell_a}^2 } ,
\end{eqnarray}
assuming real parameters and that all mixings between selectrons and
smuons are small.  Plugging these expressions into \eq{eq:amuegLLargeMu}
yields
\begin{align} \label{eq:amuegLLargeMuFinal}
&\amuegL \approx
\nonumber\\
& g_1^2 \frac{m_\mu^2}{48\pi^2} \frac{\mu\tan\beta}{M_1}
\frac{ m^2_{\tilde L_{12}} } { m_{\tilde\ell_c}^2-m_{\tilde\ell_b}^2 } 
\!\left[
\frac{ x_{1a} F_2^N(x_{1a}) - x_{1c} F_2^N(x_{1c}) }{
 m_{\tilde\ell_c}^2-m_{\tilde\ell_a}^2 } 
-\frac{ x_{1a} F_2^N(x_{1a}) - x_{1b} F_2^N(x_{1b}) }{
 m_{\tilde\ell_b}^2-m_{\tilde\ell_a}^2 }
\right] .
\end{align}

\begin{figure}%[htp]
\centering
\includegraphics{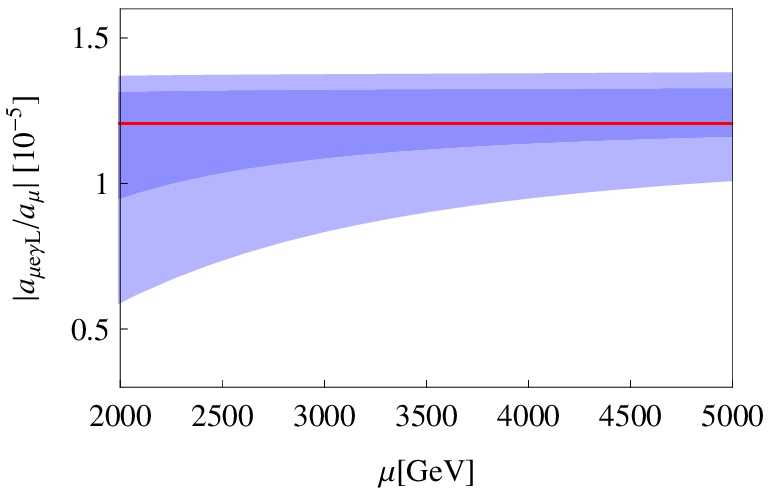}
\includegraphics{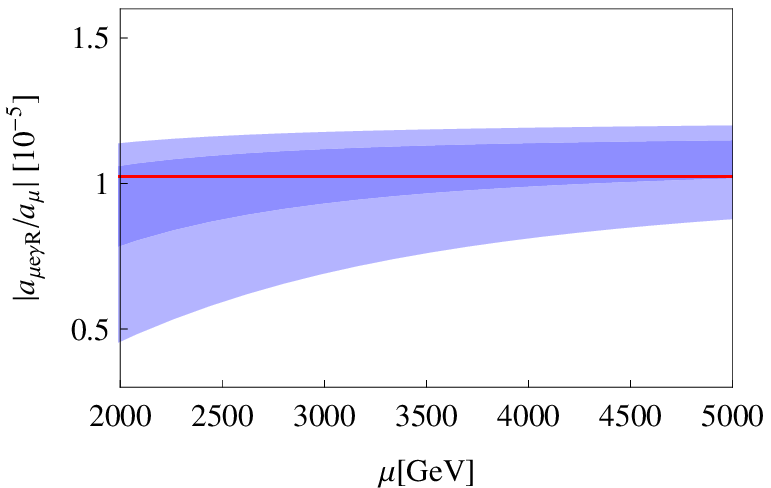}
\caption{Correlation for the case of large $\mu$, for a range
    of slepton masses. Left:
    $|\ramTLS|$ with $m_{\tilde L_{12}}= 2\GeV$, $m_{\tilde R_{12}} =0$,
    $m_{\tilde L_{11}}=470\GeV$, and all other slepton masses varied.
	The remaining parameters were set to the values given in
    column~III of \Tabref{tbl:ex_points}.
    Right: the same but left- and right-handed parameters exchanged,
    and $m_{\tilde R_{11}}$ fixed to $510\GeV$. 
    The
    light-blue shaded areas correspond to the range $[200,900]\GeV$,
    the dark-blue shaded areas to $[300,600]\GeV$, and the red lines to
    the approximations of \eq{eq:deLmUL} and \eq{eq:deRmUR},
    respectively. 
}
\label{fig:mudomexIb}
\end{figure}

The ratio $\ramTLS$ is now easily obtained from eqs.~\eqref{eq:Dmu}
and \eqref{eq:amuegLLargeMuFinal}.  We find that this ratio can be
well approximated by the numerical estimate
\begin{equation} \label{eq:deLmUL}
\left|\ramTL\right|
\approx
\frac{2}{3} \frac{|m^2_{\tilde L_{12}}|}{m^2_{\tilde\ell_a}},
\end{equation}
if all slepton masses are below a TeV.
For masses between $300\GeV$ and $600\GeV$, it deviates from the exact
result by less than 30\%, while for the wider mass range 
$[200, 900]$ GeV the difference can be up to a factor of 2.
The first plot of \Figref{fig:mudomexIb} demonstrates that the
approximation is even more accurate if we fix
$m^2_{\tilde L_{11}} = 470\GeV$ and vary only the other slepton masses.
Note that the points in the figure correspond to mass differences
$|m_{\tilde\ell_b}-m_{\tilde\ell_c}| \gtrsim 1\GeV$, which is generically
expected due to the difference between the $D$-terms of $\tilde\mu_L$
and $\tilde\mu_R$.  Fine-tuning the mass eigenvalues
could lead to extreme cancellations and thus to points outside the
colored regions in \Figref{fig:mudomexIb}. When $m^2_{\tilde R_{12}}\neq 0$ and $m^2_{\tilde L_{12}}= 0$, we can
proceed in an analogous way, and we find
\begin{equation} \label{eq:deRmUR}
\left|\ramTR\right|
\approx \frac{2}{3} \frac{|m^2_{\tilde R_{12}}|}{m^2_{\tilde\ell_a}}.
\end{equation}
This approximation is compared to the exact result in the second plot of
\Figref{fig:mudomexIb}, this time fixing $m^2_{\tilde R_{11}} = 510\GeV$.
In both plots of \Figref{fig:mudomexIb} the solid (red) lines represent
the approximations (\ref{eq:deLmUL}) and (\ref{eq:deRmUR}), while the
shaded (blue) areas originate from the random variation of parameters.

\subsection[Neutralino-smuR dominance]{Neutralino--$\boldsymbol{\tilde\mu_R}$ dominance \label{sbsc:neutsmuR}} 

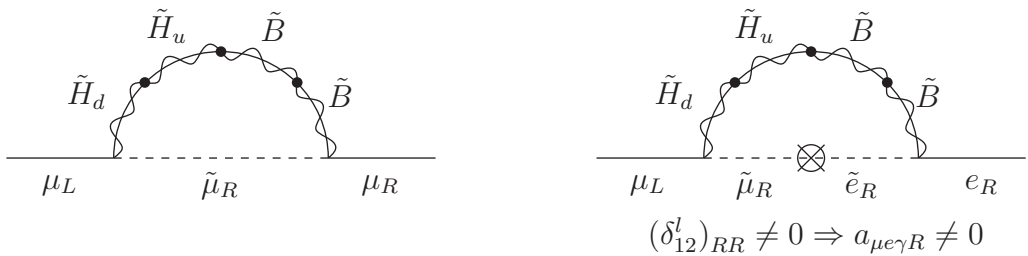
\begin{figure}
\centering
\begin{picture}(480,100)(0,0)
%%
%\Text(60,85)[1]{$ a_\mu^{L}( \tilde B, \tilde H )$}
\Line(40,30)(80,30)
\CArc(120,30)(40,360,180)
\PhotonArc(120,30)(40,0,180){3}{8}
\Vertex(120,70){2}
\Vertex(91.6,58.4){2}
\Vertex(148.4,58.4){2}
%\Line(115,75)(125,65)
%\Line(115,65)(125,75)
%\Vertex(120,70){2}
%\DashArrowLine(80,30)(160,30){3}
\DashLine(80,30)(160,30){3}
\Line(160,30)(200,30)
\Text(60,20)[1]{$\mu_L$}
\Text(120,20)[1]{$\tilde\mu_R$}
\Text(180,20)[1]{$\mu_R$}
%\Line(175,35)(185,25)
%\Line(175,25)(185,35)
%\Text(120,85)[1]{$\tilde B$}
\Line(260,30)(300,30)
\CArc(340,30)(40,360,180)
\PhotonArc(340,30)(40,0,180){3}{8}
%\Text(340,85)[1]{$\tilde B$}
\Text(320,80)[1]{$\tilde H_u$}
\Text(360,80)[1]{$\tilde B$}
\Text(290,55)[1]{$\tilde H_d$}
\Text(385,55)[1]{$\tilde B$}
%\DashArrowLine(300,30)(380,30){3}
\DashLine(300,30)(380,30){3}
\Line(380,30)(420,30)
\Line(335,35)(345,25)
\Line(335,25)(345,35)
\Text(280,20)[1]{$\mu_L$}
\Text(320,20)[1]{$\tilde \mu_R$}
\Text(360,20)[1]{$\tilde e_R$}
\Text(405,20)[1]{$e_R$}
\Text(100,80)[1]{$\tilde H_u$}
\Text(140,80)[1]{$\tilde B$}
\Text(70,55)[1]{$\tilde H_d$}
\Text(165,55)[1]{$\tilde B$}
\Vertex(340,70){2}
\Vertex(311.6,58.4){2}
\Vertex(368.4,58.4){2}
%\Line(335,75)(345,65)
%\Line(335,65)(345,75)
\CArc(340,30)(5,0,360)
\Text(343,1)[1]{$\deltaRR \neq 0 \Rightarrow \amuegR \neq 0$}
\end{picture}
\caption{Diagrams corresponding to the leading contributions to $a_\mu$ and $\amuegR$ in the case of $\tilde\mu_R$ dominance.}
\label{eq:smuRneut}
\end{figure}

We consider the contributions from the diagrams of \Figref{eq:smuRneut},
which involve the right-handed smuon $\tilde{\mu}_R$.  They dominate if
the spectrum satisfies
\begin{equation}
\label{eq:hier_lightsmuR}
M_1,\; m_{\tilde\mu_R} ,\; m_{\tilde e_R} < M_2, \; |\mu| \ll \; m_{\tilde\mu_L} ,\; m_{\tilde e_L} ,
\end{equation}
where the hierarchy $M_1 < M_2, |\mu|$ ensures a bino-like lightest neutralino.
Note that  $\mu<0$ in order to have a positive $a_\mu$
\cite{Stockinger:2006zn,Grothaus:2012js}. The contributions from the diagrams of
Figs.~\ref{eq:spCh_diagrams} and~\ref{eq:spNeut_diagrams} are
suppressed due to the large values of the left-handed slepton masses.
\paragraph{Interplay of contributions to $\boldsymbol{a_\mu}$} 
From the left diagram of \Figref{eq:smuRneut} we can then estimate $a_\mu$
as%
\footnote{We use
$N_{11} N_{13} \approx \frac{M_Z \sin\theta_W \sin\beta}{\mu}$,
which is a good approximation if \eq{eq:hier_lightsmuR} holds and
in addition
$|\mu| \gg M_Z$ and $\tan\beta \gtrsim 30$ \cite{Gunion:1987yh}.
}
\begin{eqnarray} \label{eq:Damu_muRLarge}
a_\mu &\approx&
-g_1 g_2 \frac{m_\mu^2}{48\pi^2 M_W \cos\beta} \frac{1}{M_1} \,
\text{Re}[N_{11} N_{13}] \, x_{1a} F_2^N(x_{1a})
\nonumber\\
&\approx& -g_1^2 \frac{m_\mu^2}{48\pi^2}
\frac{\tan\beta}{M_1 \mu} \, x_{1a} F_2^N(x_{1a}) ,
\end{eqnarray}
where we have labeled the state which is mostly $\tilde\mu_R$ as $\tilde\ell_a$.
Numerically, we can summarize the behavior of $a_\mu$ for this case as
\begin{equation}
a_\mu \approx 3 \times 10^{-9} \, x_{1a} F_2^N(x_{1a})
\left( \frac{\tan\beta}{50} \right)
\left( \frac{500\GeV}{-\mu} \right)
\left( \frac{100\GeV}{M_1} \right) .
\end{equation}
{For the set of parameters in column IV of \Tabref{tbl:ex_points},
the $\tilde\chi^0$--$\tilde\mu_R$ contribution dominates and the value
of $a_\mu$ is within the allowed $1\sigma$ region if $m_{\tilde\mu_R} \lesssim 90\GeV$, within the $2\sigma$ region if $m_{\tilde\mu_R} \lesssim 130\GeV$ and within
the $3\sigma$ region if $m_{\tilde\mu_R} \lesssim 220\GeV$.}
 {A benchmark parameter point with this behavior
 {and a value of $a_\mu$ in the allowed $2\sigma$ range}
was also defined and discussed in Ref.\ \cite{Fargnoli:2013zda}.}

\begin{figure}%[htp]
\centering
\includegraphics{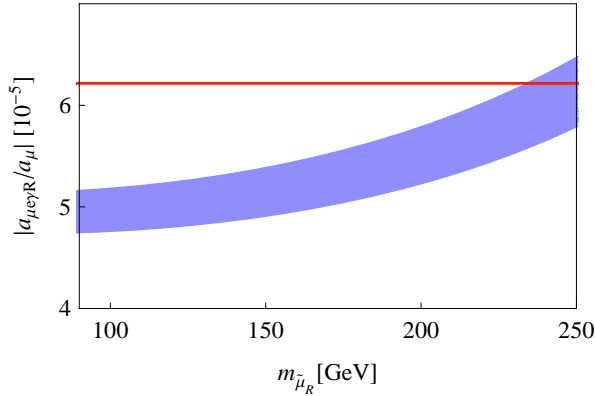}
\caption{Correlation for the case of $\tilde{\mu}_R$ dominance.  The
  band shows the ratio $|\ramTRS|$ as a function of
$m_{\tilde\mu_R}$, for a random variation of $\mu$ in
the range $[-550,-650]\GeV$ and $M_2$ in the range $[100,900]\GeV$. The
rest of the parameters are as in column IV of
\Tabref{tbl:ex_points}. The approximation  \eq{eq:approx_smR}
 is represented by the horizontal line.
\label{fig:ratamuegoasm_smR}}
\end{figure}

\paragraph{Interplay of contributions to $\boldsymbol{a_{\mu e \gamma}}$}
For this case, the relevant flavor-violating amplitude is
\begin{equation}
\amuegR \approx
-g_1 g_2 \frac{m_\mu^2}{48\pi^2 M_W \cos\beta} \frac{1}{M_1} \,
N_{11}^* N_{13}^* \, \sum_m K_{m4}^* K_{m5} \, x_{1m} F^N_2(x_{1m}) ,
\end{equation}
where sizable contributions come only from the slepton mass eigenstates
containing mainly $\tilde\mu_R$ and $\tilde e_R$, which we label as
$\tilde\ell_a$ and $\tilde\ell_b$, respectively.
The mixing of the two states is given by
$|K_{a5} K_{a4}| \approx
|m^2_{\tilde R_{12}} / (m_{\tilde\ell_a}^2-m_{\tilde\ell_b}^2)|$
and $K_{b5} K_{b4} \approx -K_{a5} K_{a4}$
for real parameters.
Using the same approximation for $N_{11} N_{13}$ as before, we find
\begin{equation}
|\amuegR| \approx
g_1^2 \frac{m_\mu^2}{48\pi^2} \frac{\tan\beta}{M_1 \, |\mu|} \left|
m^2_{\tilde R_{12}} 
\frac{ x_{1a} F_2^N(x_{1a}) - x_{1b} F_2^N(x_{1b}) }{ m_{\tilde\ell_a}^2-m_{\tilde\ell_b}^2 } 
\right| .
\end{equation}
Taking into account \eq{eq:Damu_muRLarge}, $|\ramTRS|$ becomes 
\begin{equation} \label{eq:approx_smR}
\left|\ramTR\right| \approx
\left| 
\frac{ m^2_{\tilde R_{12}} }{ m_{\tilde\ell_a}^2-m_{\tilde\ell_b}^2 } 
\left( 1 - \frac{x_{1b} F_2^N(x_{1b})}{x_{1a} F_2^N(x_{1a})} \right)
\right| \sim
\frac{|m^2_{\tilde R_{12}}|}{m^2_{\tilde\ell_b}}.
\end{equation}
In \Figref{fig:ratamuegoasm_smR} we show the numerical results for
$|\ramTRS|$ as a blue band, taking into account the chargino contribution
as well. The parameters are set to the values of
\Tabref{tbl:ex_points}, column IV; in particular, we varied $\mu$ in the range $[-550,-650]\GeV$ and $M_2$ in
the range $[100,900]\GeV$.  
For $\mu<-650\GeV$,
$a_\mu$ leaves the allowed $3\sigma$ region for the
considered mass spectrum.  Note that as long as the left-handed slepton
masses are kept above $2\TeV$, there is no change of the behavior
presented for this scenario. 
The horizontal line in \Figref{fig:ratamuegoasm_smR} represents the approximation of \eq{eq:approx_smR}.
The width of the band in the figure and its limited variation with
$m_{\tilde\mu_R}$ allow us to conclude that the case of
neutralino-$\tilde\mu_R$ domination features a strong correlation
between $a_\mu$ and \BR.  

\section{Discussion and main results \label{sec:discussions}}

\subsection[Correlation between amu and BR for similar supersymmetric masses]{Correlation between $a_\mu$ and $\BR$ for similar supersymmetric masses}

Our study started with the following basic question:
Assuming supersymmetric parameters that reproduce the observed value of
$\amuG$, can we predict the amplitude $\amuegL$ (or $\amuegR$)? Evidently this
amplitude depends on the flavor-violating parameter $m^2_{\tilde L_{12}}$
(or $m^2_{\tilde R_{12}}$),
so in the best possible case it would be proportional to $a_\mu$ times
the dimensionless ratio of the flavor-violating parameter and some other SUSY mass $m_{\tilde{p}}^2$. I.e.\ in the best possible case we could write
\begin{align}
\left|\ramTL\right| & \approx
f\frac{m^2_{\tilde L_{12}}}{m^2_{\tilde{p}}}
\end{align}
with a constant $f$. We found several parameter regions in which such correlations hold for $\amuegL$ or $\amuegR$,
but the proportionality constant $f$ and the appropriate mass ratio are specific for each region.

In \Secref{sec:SimilarMasses} we found that for similar masses of the
supersymmetric particles involved in $\amuG$ and $\amuegG$, we can indeed determine the order of magnitude of $\amuegL/\amuG$,
employing as mass ratio the commonly used quantity $\deltaLL$ 
defined in \eq{eq:classicfvp}.
The correlation is rather weak, however, and we also found that significant cancellations
among different diagrams contributing to both processes are typical.

\begin{figure}
\centering
\includegraphics{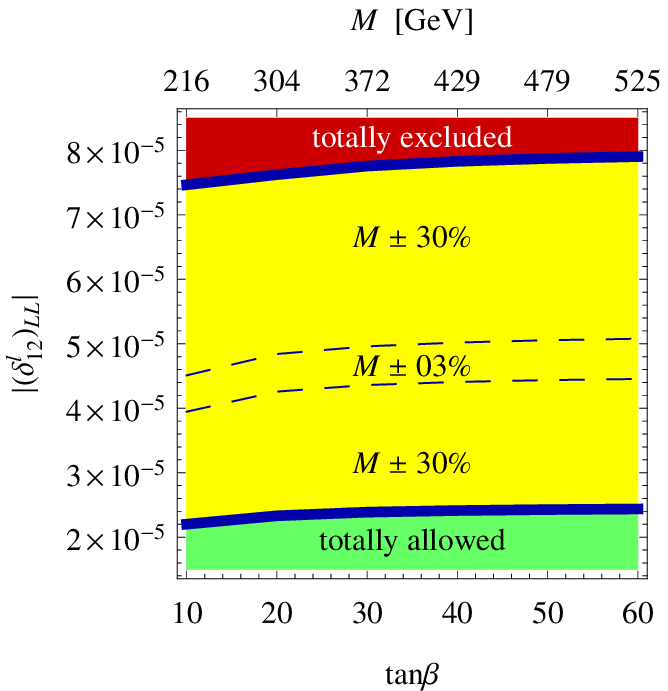}
\includegraphics{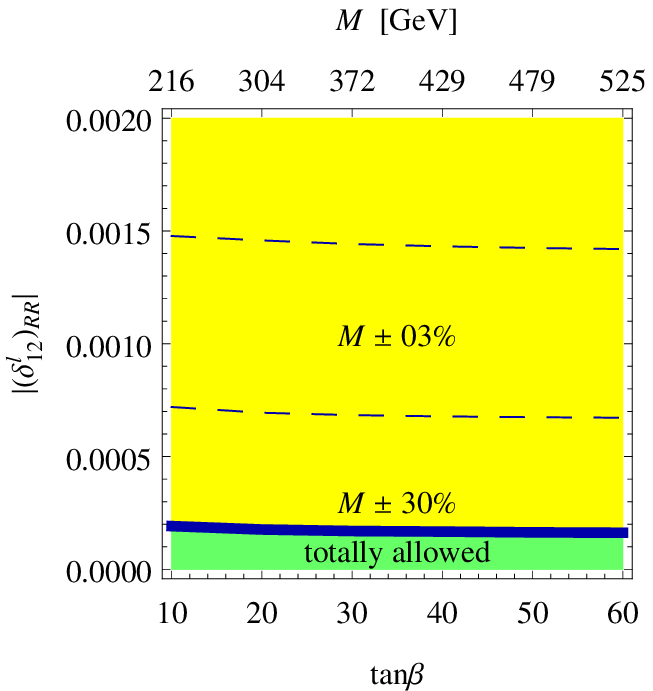}
\caption{Strong and weak bounds on
$\deltaLL$ and $\deltaRR$
for two different mass ranges: $(1\pm 0.3)M$ and $(1\pm 0.03)M$.
See text for details.}
\label{fig:boundsondeltas}
\end{figure}

Nevertheless, using the MEG limit on \BR~\cite{Adam:2013mnn} we can put
bounds on the flavor-violating parameters under the assumptions that $a_\mu$ is explained by SUSY and all relevant SUSY masses are similar.
In the left panel of \Figref{fig:boundsondeltas} we plot bounds on $\deltaLL$ as a function of $\tan\beta$. The bounds are obtained from a random scan as follows.
For each value of  $\tan\beta$, the value of the generic parameter $M$ is chosen such that $a_\mu$ agrees with its central experimental counterpart 
in \eq{eq:Dmuallw_exp} when
\begin{equation}
\mu = M_1 = M_2 = m_{\tilde L_{11}} = m_{\tilde L_{22}} = m_{\tilde R_{11}}  = m_{\tilde R_{22}} = M.
\end{equation}
Afterwards, random SUSY mass spectra are generated by
varying the seven mass parameters above within the
interval $[0.7 M, 1.3 M]$, imposing the conditions that
(a) the LSP is a neutralino and (b) $a_\mu$ falls into the
$1\sigma$ region given in \eq{eq:Dmuallw_exp}.
From
the generated spectra we derive three regions and two
corresponding bounds on $\deltaLL$. The top-most (red) region in the
left panel of \Figref{fig:boundsondeltas} is
``totally excluded'', i.e.\ $\deltaLL$ is so large that the MEG limit
is violated by all generated mass spectra. The
lower-most (green) region is ``totally allowed'', i.e.\ the MEG limit
is never violated.  In the yellow region in between
some (but not all) spectra satisfy the MEG limit.
Thus, the upper, weaker bound
delimiting the red region is conservative and must be satisfied
(under the above assumptions).
The yellow region is allowed by the above assumptions, but with more
information on the SUSY masses from either experiment 
or theoretical models, the bound on $\deltaLL$ might go down as low as the
lower, stronger bound delimiting the green region.

In order to see how the size of the mass range affects the bounds, we
also investigate the smaller interval $(1 \pm 0.03) \times M$.
The result is a narrower band limited by the dashed boundaries in the
plot, which are analogous to the thick blue lines.

In addition to what is shown in the plot, we calculated bounds from a
restricted set of spectra,
imposing $m_{\tilde L_{11}} = m_{\tilde L_{22}}$ and $m_{\tilde R_{11}} = m_{\tilde R_{22}}$.
In comparison to the general case, the additional degeneracy conditions result in a slightly relaxed strong bound.
One can understand this from the fact that the conditions restrict the
spread of $\mathrm{BR}(\mu \rightarrow e\gamma)$ for a fixed mass
insertion. By the same token, the weak bound becomes tighter, albeit
only by a tiny margin.

In the right plot of \Figref{fig:boundsondeltas}, we consider flavor
violation in the right-handed slepton sector. The weak bound on
$\deltaRR$ is not displayed.  We omit it since the calculation is based
on the approximation $\BR \propto |\deltaRR|^2$.  Naively applying this
relation would result in bounds of $\mathcal{O}(1)$.  For so large
values of $\deltaRR$, the proportionality approximation is not valid.
In any case, the weak bound is expected to lie
far outside the plotted vertical range.
For such a large mass insertion,
it might not be $\BR$
but the smallest slepton mass eigenvalue
that determines the limit on $\deltaRR$.
For $|(\delta^l_{12})_{RR}| \ll 1$,
one can always find a set of mass parameters
within $[0.7M, 1.3M]$ such that
different contributions to $\mu \rightarrow e\gamma$ cancel
\cite{Masina:2002mv,Paradisi:2005fk},
resulting in a decay rate below the MEG limit.
The $3\%$ range, however, is too narrow for a sufficient
cancellation.  This leads to the upper dashed curve, which depicts the
weak bound for this small mass range.

After considering the case of similar superparticle masses, we turned to
the next step, asking under which conditions a strong correlation arises.
The answer is that there should be just one kind of diagram (where only one kind of supersymmetric particle mediates the contribution to both processes) and so we explored cases where either charginos or neutralinos could dominate.

\subsection{Specific results for the case of chargino dominance}

At the end of \Secref{sec:chargino dominance} we have made a
comparison between the accuracy of the correlations in cases where the charginos
dominate both $\amuG$ and $\amuegG$ and the general correlation in the
case of \Secref{sec:SimilarMasses}. We have also pointed out for which cases
the correlations become particularly useful. In summary, depending on
the hierarchy of the sneutrino and chargino masses, the different
approximations given in \Tabref{tbl:ch_correlations}  predict the ratio
of the lighter chargino contributions to $\amuegL$ and $a_\mu$, i.e.\
the ratio $\amuegII/\amuII$, within a factor $1.5$, except for case~IV\@. In cases I (for $M_2 \gtrsim 1\TeV$), II, III, V, VI, and IX, these
approximations can even be used as a reliable substitute for the full value of $\amuegL/a_\mu$ stemming from all contributions.

\begin{figure}%[htp]
\centering
\includegraphics{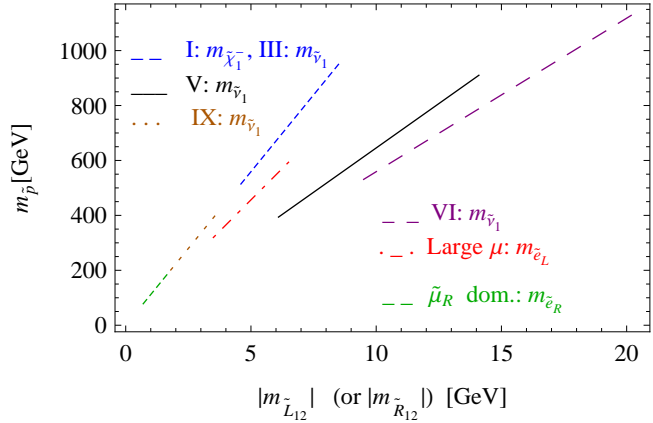}
\caption{Lower bounds on the masses of the supersymmetric particles that
drive the ratio $\ramTLS$ as a function of the off-diagonal mass
$m_{\tilde{L}_{12}}$ or $m_{\tilde{R}_{12}}$ for $\tan\beta=50$. Cases I, III, V, VI and IX refer to the cases of chargino dominance, \Secref{sec:chargino
dominance}, for which $m_{\tilde{L}_{12}}$ is relevant.  For the case of
large $\mu$ (where the neutralino contribution dominates),
\Secref{sec:LargeMu}, $m_{\tilde{L}_{12}}$ sets the bound for $m_{\tilde e_L}$
while $m_{\tilde{R}_{12}}$  sets the bound for $m_{\tilde e_R}$. The
case of $\tilde\mu_R$ dominance, \Secref{sbsc:neutsmuR}, is only sensitive to  $m_{\tilde{R}_{12}}$.
}
\label{fig:lowerboundsmChdom}
\end{figure}

Again, we can in turn use the obtained correlations to set bounds on the
relevant parameter combinations appearing in $\ramTLS$. In order to highlight the structure of the bounds and the differences in the different parameter regions we do not carry out scans in parameter space but we
 use the experimental information on $a_\mu$ from \eq{eq:Dmuallw_exp}
and require $\BR$ to satisfy the bound (\ref{expval:BRemugamma}). Then we obtain
\begin{equation} \label{eq:bdondL12}
\left|\ramTL\right| < 8 \times 10^{-5} \,
\left| \frac{287 \times 10^{-11}}{a_\mu} \right| ,
\end{equation}
where we have omitted $|\amuegR|^2$ in the expression for \BR\ since it is
subdominant. In all cases of \Tabref{tbl:ch_correlations} in which the
left-hand side can be approximated reliably as a simple mass ratio this
bound translates into a lower bound on the corresponding mass
$m_{\tilde p}$ of the supersymmetric particle driving the correlation
between $a_\mu$ and $\BR$; for example, for case I this is
$m_{\tilde\chi^\pm_1}$. The bounds are
functions of the flavor-violating parameter $m_{\tilde L_{12}}$. The results are shown in
\Figref{fig:lowerboundsmChdom}, where we have plotted only those cases
of \Tabref{tbl:ch_correlations} for which a reliable bound can be
extracted.
For each case we have restricted $m_{\tilde p}$ to the range where the
approximation for $\amuegL/a_\mu$ is valid within a factor $1.5$
and where $a_\mu$ can be made to lie in the favored $2\sigma$ region
by choosing the superparticle masses appropriately.
 We can see that these
requirements restrict considerably the scale $m_{\tilde p}$.
 In the figure $\tan\beta$ is fixed to 50; lowering (increasing) it
would allow to loosen the lower (upper) bounds on $m_{\tilde p}$.
Note that among the cases mentioned here, we have left out
case II because the useful approximation involves logarithms and so it is not possible to set a bound on the mass of an individual particle.

\subsection{Specific results for cases with neutralino dominance}

We identified two particular regions of parameter space, also represented by the benchmark points 3 and 4 of Refs.\ \cite{Fargnoli:2013zda,Fargnoli:2013zia}, where neutralino contributions dominate and  $\amuG$ and $\amuegG$ are strongly correlated.

If $\mu$ is very large and the bino mass $M_1$ sufficiently small, the exchange of the lightest, bino-like neutralino can
become the dominant contribution. Contrary to what happens in the case
of chargino dominance, for the neutralino case the contributions from
$\amuegR$ can be relevant, since the lightest neutralino for this case is
bino-like (see \Figref{eq:spNeut_diagrams}). Hence both left- and
right-handed slepton contributions to $\amuegG$ and to $\amuG$ become important.

  For this case, we have found in \Secref{sec:LargeMu}  that indeed a single diagram dominates and
there is a strong correlation between $a_\mu$ and $\BR$. 
Hence we analyzed  separately the correlations when either
left-handed or right-handed charged sleptons dominate $\amuG$ (and thus also $\amuegG$). The correlations lead to relations of the type
\[
\left|\ramTL\right| \approx \frac{2}{3} \frac{|m^2_{\tilde L_{12}}|}{m^2_{\tilde\ell_a}},\quad
\left|\ramTR\right| \approx \frac{2}{3} \frac{|m^2_{\tilde R_{12}}|}{m^2_{\tilde\ell_a}},
\]
where $\tilde\ell_a \approx \tilde e_L$ and $\tilde\ell_a \approx \tilde
e_R$, respectively. In this sense, just as in the case of chargino
dominance, we can obtain reliable lower bounds on the mass of the supersymmetric particle driving the correlation between $a_\mu$ and $\BR$. We plot the results  in \Figref{fig:lowerboundsmChdom}, where we can also compare to the bounds in the case of the chargino dominance.

Another case of neutralino dominance for which we have found a strong correlation is the case analyzed in 
\Secref{sbsc:neutsmuR}, where $a_\mu$ 
is dominated by diagrams involving the right-handed smuon and the lightest neutralino. 
The conditions for this dominance are that
(a)~left-handed slepton masses be much more bigger than $|\mu|$ and
(b) $M_1, m_{\tilde\mu_R}, m_{\tilde e_R} < M_2$. We find that
\[
\left|\ramTR\right| \sim \frac{|m^2_{\tilde R_{12}}|}{m^2_{\tilde\ell_a}},
\]
where now obviously $\tilde\ell_a \approx \tilde e_R$. The lower bound
on the mass of $\tilde\ell_a$, as a function of $|m^2_{\tilde R_{12}}|$,
is also visualized in \Figref{fig:lowerboundsmChdom}.

\subsection{Comparison of different scenarios}

\begin{figure}%[htp]
\centering
\includegraphics[width=7.4cm]{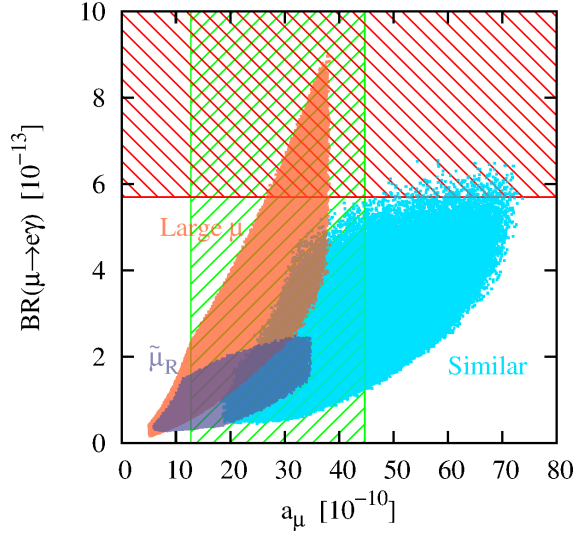}
\caption{Comparison of the large $\mu$ and the $\tilde\mu_R$ dominance scenarios
with the case of similar supersymmetric masses. See text for details.}
\label{fig:islandsplot}
\end{figure}

Figure \ref{fig:islandsplot} summarizes the correlations found in the three most important cases.  
The plot displays three ``islands'': (a) the similar supersymmetric masses
case, \Secref{sec:SimilarMasses}, (b) the case of large $\mu$, and (c) $\tilde\mu_R$ dominance,
for $\tan \beta = 50$ and $\deltaLL= \deltaRR=  2\times 10^{-5}$.  The mass ranges are as
follows: For scenario (a) all the seven mass
parameters vary between $300\GeV$ and $600\GeV$. For case (b),
$M_1$ varies as in (a), while the slepton masses vary 
between $450\GeV$ and $900\GeV$, and $\mu = M_2$ are fixed at $4\TeV$.
Finally, for scenario (c) we have chosen 
$M_1 \in [100,150]\GeV$,
$M_2 \in [500,2000]\GeV$,
$-\mu \in [300,600]\GeV$,
$m_{\tilde\ell_R} \in [100,200]\GeV$ and 
$m_{\tilde\ell_L} = 3\TeV$.
The figure confirms the existence of correlations, and it allows to easily look up the expected results for each parameter region. In detail, let us explain why we observe a slightly larger variation of \BR\ for fixed $a_\mu$ than found
earlier, for example in \Figref{fig:mudomexIb}. Earlier we fixed the very parameter
ratio driving the correlation between $a_\mu$ and $\BR$, for example
$m^2_{\tilde L_{12}}/m^2_{\tilde\ell_a}$.
In \Figref{fig:islandsplot} we fixed
$\deltaLL \approx m^2_{\tilde L_{12}}/m^2_{\tilde\ell_a} \times m_{\tilde e_L}/m_{\tilde\mu_L}$,
i.e.\ the decisive parameter multiplied by a factor varying between
$0.5$ and $2$ if we vary the slepton masses between $450\GeV$ and
$900\GeV$. 
This illustrates a certain limitation of the correlations: Even if we
consider just a single diagram, varying all SUSY masses by a factor of
$2$ can still change the branching ratio by an order of magnitude,
unless we fix the ``correct'' flavor-violating parameter, which varies
from diagram to diagram.

\begin{figure}%[htp]
\centering
\includegraphics[width=7.4cm]{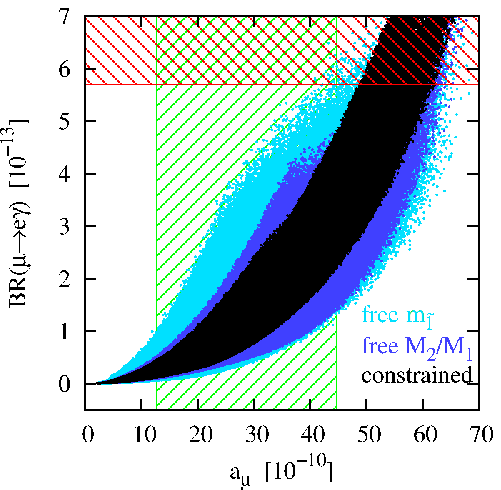}
\quad
\includegraphics[width=7.4cm]{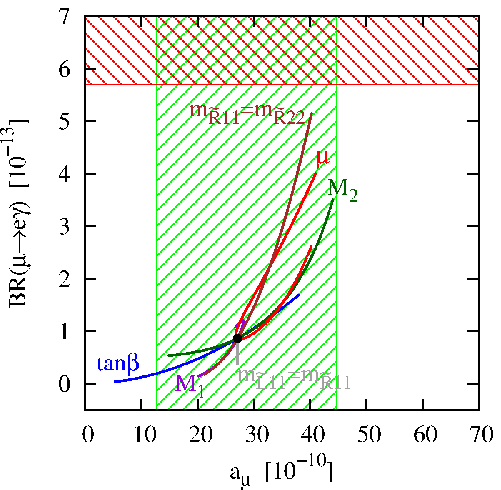}
\caption{Left: comparison of the results found in \cite{Isidori:2007jw} to our
similar supersymmetric mass scenario.  The parameter ranges are $300\GeV
\le m_{\tilde L_{11}}, m_{\tilde L_{22}}, m_{\tilde R_{11}}, m_{\tilde
R_{22}} \le 600\GeV$, $100\GeV \le M_1 \le 500\GeV$, $200\GeV \le M_2
\le 1000\GeV$, $500\GeV \le \mu \le 1000\GeV$, $10 \le \tan\beta \le 50$.
The black band obeys the constraints $M_2 = 2 M_1$, $m_{\tilde L_{11}}= m_{\tilde L_{22}}= m_{\tilde R_{11}}= m_{\tilde R_{22}}$.
In the dark blue (dark grey) region, $M_1$ and $M_2$ are independent of each other.
In the light blue (light grey) region, the four $m_{\tilde {L}_{ii}/\tilde{R}_{ii}}$ are independent of one another.
Right: drift of a point due to the change of each variable,
around the black dot with
$\tan\beta = 50$,
$M_1 = 300\GeV$, $M_2 = 600\GeV$, $\mu = 750\GeV$,
$m_{\tilde L_{11}}= m_{\tilde L_{22}}= m_{\tilde R_{11}}= m_{\tilde R_{22}} = 380\GeV$.
}
\label{fig:comparisonplot}
\end{figure}

Finally, let us compare our results to previous studies in the
literature where stronger correlations were found. In particular, Fig.~6
of \cite{Isidori:2007jw}
shows a narrower band than the one for the case of similar
supersymmetric masses
plotted in our
\Figref{fig:islandsplot}.
To clarify, in {the left panel of} \Figref{fig:comparisonplot} we reproduce our results for
the similar masses scenario together with the results of Fig.~6 of
\cite{Isidori:2007jw} using the same constraints as there, except $A_U$,
which we set to zero, and $\deltaLL$, which we set to $2\times 10^{-5}$.
Specifically, in \cite{Isidori:2007jw} the additional constraints with respect to ours are $M_2\approx 2 M_1$ and 
$m_{\tilde L_{11}}= m_{\tilde L_{22}}= m_{\tilde R_{11}}= m_{\tilde R_{22}}$.
The resulting band is the black (darkest) one in our
\Figref{fig:comparisonplot}. Indeed it agrees with the corresponding one in Ref.\ \cite{Isidori:2007jw}. 
Relaxing the constraint on the gaugino
masses produces a wider, dark blue (dark grey) band, and relaxing the constraints on the slepton masses produces the light blue (light grey) band.
This shows that the correlation found in \cite{Isidori:2007jw} is
stronger than the one we find (for our similar supersymmetric mass scenario) because of the constraints on the slepton and gaugino masses.

Furthermore, the right panel of \Figref{fig:comparisonplot}
  displays an anatomy of the decorrelation effect arising from the
  variation of each parameter.  First, we pick up a representative
  point, marked by the black dot, out of the black band in the left
  panel.  Then, we vary each of the dimensionful parameters shown in
  the plot by a factor of 10 around the dot, for instance,
  $240 \GeV \le \mu \le 2400 \GeV$.
  The range of $\tan\beta$ is from 10 to 70.
  One might regard the $\tan\beta$ curve as depicting
  the principal correlation that guides the black band, as
  both $a_\mu$ and $\amuegL$ are approximately proportional to $\tan\beta$.
  The other curves are then deviations from this principal correlation.
  Among them,
  the $m_{\tilde L_{11}}=m_{\tilde R_{11}}$ line should be the
  easiest to understand:
  It shows simply that $a_\mu$ remains fixed while
  $\BR$ decreases as the selectron masses increase.
  One can notice in the left panel that
  relaxing the equality of the four diagonal slepton masses
  causes a greater spread than
  unfixing the gaugino mass ratio.
  In the right plot,
  the slepton mass splits are further divided into two classes:
  the intergeneration split and the left-right split.
  Comparing the $m_{\tilde L_{11}}=m_{\tilde R_{11}}$
  and the $m_{\tilde R_{11}}=m_{\tilde R_{22}}$ curves,
  one finds that
  the left-right split causes
  a larger deviation from the principal correlation.

In conclusion, the essential goal of the present work was to 
characterize the sources of the possible correlation between
$a_\mu$ and \BR. We discerned that in the case of similar SUSY masses
entering both observables cancellations are typical and hence the
correlation is rather weak; we also identified some cases where the correlation is
strong. In these cases we derived bounds on the flavor-violating parameters
under various assumptions (\Figref{fig:boundsondeltas}) and conversely
bounds on the masses of supersymmetric particles, as a function of
only one flavor-violating 
parameter (\Figref{fig:lowerboundsmChdom}). 
The ultimate application of this kind of analyses would be attained if $\BR$ could eventually be measured, because that would 
give definite indications for the required values of all relevant mass
parameters.
Finally,
\Figref{fig:islandsplot} and \Figref{fig:comparisonplot} can be
regarded as a summary, and as an update and generalization of
Ref.\ \cite{Isidori:2007jw}. The looser the assumptions in particular
on slepton masses, the weaker
the correlations, but there are interesting further parameter
space islands with strong mass hierarchies, where correlations exist.

\subsection*{Acknowledgments}
This work was supported by the German Research Foundation (DFG)
via the Junior Research Group ``SUSY Phenomenology'' and an SFB
Fellowship for L.~Velasco-Sevilla within the Collaborative Research Center 676
``Particles, Strings and the Early Universe''.  
J.P. acknowledges support from the MEC and FEDER (EC) Grants
FPA2011--23596 and the Generalitat Valenciana under grant PROMETEOII/2013/017.
We thank Zackaria Chacko for discussions.

\providecommand{\href}[2]{#2}\begingroup\raggedright\endgroup

\end{document}